\documentclass[journal]{IEEEtran}
\usepackage{amssymb,amsmath,amsfonts,bm,epsfig,graphicx,theorem,latexsym}
\usepackage{rotating,setspace,latexsym,epsf,color,subfigure}
\usepackage{cite,times,authblk,epstopdf}
\usepackage{multicol}%
\usepackage{verbatim}%
\usepackage{enumerate}%------------------------
\usepackage{eucal}
\usepackage{caption}
\usepackage{stmaryrd,bbm,subfigure,units}

\newtheorem{theorem}{Theorem}

\newtheorem{corollary}{Corollary}

\newtheorem{lemma}{Lemma}

\newcounter{remark}
\newenvironment{remark}[1][]{\refstepcounter{remark}\medskip\par\noindent{\bf Remark~\theremark\,~}\,#1}{{\mbox{\,$\blacksquare$}\medskip\par}}

\newenvironment{Proof}[1]{\medskip\par\noindent{\bf Proof:\,}\,#1}{{\mbox{\,$\blacksquare$}\medskip\par}}

\makeatletter

\allowdisplaybreaks[1]

\newcommand{\bX}{\mathbf{X}}
\newcommand{\bx}{\mathbf{x}}

\newcommand{\bZ}{\mathbf{Z}}

\newcommand{\bd}{\mathbf{d}}
\newcommand{\bW}{\mathbf{W}}
\newcommand{\bK}{\mathbf{K}}

\newcommand{\setA}{\mathcal{A}}
\newcommand{\setB}{\mathcal{B}}
\newcommand{\setC}{\mathcal{C}}

\newcommand{\setW}{\mathcal{W}}
\newcommand{\setZ}{\mathcal{Z}}
\newcommand{\limitn}{{\underset{n \rightarrow\infty}\lim}}

\title{Coded Caching in the Presence of a Wire and a Cache Tapping Adversary of Type II}

\author{Mohamed~Nafea{\dag},~\IEEEmembership{Member,~IEEE,}
        and Aylin~Yener{\ddag},~\IEEEmembership{Fellow,~IEEE}
\thanks{Manuscript received August 15, 2020; revised December 04, 2020; accepted January 14, 2021. This work was supported in part by NSF Grant CCF 2105872. This article was presented in part at the 2018 IEEE Information Theory Workshop \cite{nafea2018caching1} and 2018 Allerton Conference on Communication, Control, and Computing \cite{nafea2018caching2}.

Mohamed Nafea is with the Electrical and Computer Engineering Department, University of Detroit Mercy, Detroit MI 48221, USA (e-mail: nafeamo@udmercy.edu).

Aylin Yener is with the Departments of Electrical and Computer Engineering, Integrated Systems Engineering and Computer Science and Engineering, The Ohio State University, Columbus OH 43210, USA (e-mail: yener@ece.osu.edu).

Communicated by Rafael F. Schaefer, Guest Editor for the special issue on privacy and security of information systems.}}

\begin{document}

\IEEEoverridecommandlockouts

\maketitle

% ------------------------------ ABSTRACT -----------------------------
\begin{abstract}
This paper introduces the notion of {\it{cache-tapping}} into the information theoretic models of coded caching. The wiretap channel II in the presence of multiple receivers equipped with fixed-size cache memories, and an adversary which selects symbols to tap into from cache placement and/or delivery is introduced. The legitimate terminals know neither whether placement, delivery, or both are tapped, nor the positions in which they are tapped. Only the size of the overall tapped set is known. For two receivers and two files, the strong secrecy capacity-- the maximum achievable file rate while keeping the overall library strongly secure-- is identified. Lower and upper bounds on the strong secrecy file rate are derived when the library has more than two files. Achievability relies on a code design which combines wiretap coding, security embedding codes, one-time pad keys, and coded caching. A genie-aided upper bound, in which the transmitter is provided with user demands before placement, establishes the converse for the two-files case. For more than two files, the upper bound is constructed by three successive channel transformations. Our results establish provable security guarantees against a powerful adversary which optimizes its tapping over both phases of communication in a cache-aided system.
\end{abstract}

\section{Introduction}\label{Int}
Caching aims to reduce network traffic congestion by pro-actively storing partial contents at the cache memories of end users during off-peak times, providing local caching gain \cite{dowdy1982comparative,almeroth1996use,borst2010distributed}. Seminal work in \cite{maddah2014fundamental} has shown that, careful design of cache contents in a multi-receiver setting allows the transmitter to send delivery transmissions that are simultaneously useful for many users, providing a further gain termed as the {\it{global caching gain}}. This gain depends on the aggregate cache memory of the network and demonstrates the ability of coding over delivery transmission and/or cache contents.

In numerous works to date, coded caching has been studied under various modeling assumptions and network configurations, including decentralized caching\cite{maddah2015decentralized}, non-uniform demands\cite{niesen2017coded}, more users than files\cite{wan2016caching,sahraei2016k,amiri2016coded}, heterogeneous cache sizes\cite{ibrahim2019coded}, improved bounds\cite{lim2017information,amiri2017fundamental,yu2018exact}, hierarchical caching\cite{karamchandani2016hierarchical},  interference networks\cite{maddah2015cache,hachem2016degrees,naderializadeh2017fundamental}, combination networks\cite{ji2015fundamental,zewail2017combination}, device-to-device communication\cite{ji2016fundamental,ibrahim2020device}, coded placement\cite{zhang2018fundamental,ibrahim2018benefits}, delivery over noisy channels \cite{bidokhti2016noisy,ghorbel2016content,amiri2017cache,zhang2017fundamental,bidokhti2017benefits}. 

Coded caching with security guarantees has been studied in \cite{sengupta2015fundamental,awan2015fundamental,ravindrakumar2017private,zewail2020device,zewail2017combination,zewail2019secure, kamel2018secrecy,zewail2018wiretap,zewail2019untrusted}. These, as they pertain to an external adversary, i.e., a wiretapper, assume secure cache placement; the adversary cannot tap into the cache nor the communication which performs cache placement. At the other extreme, if cache placement were to be public, i.e., if the adversary has perfect access to the cache contents, it follows from \cite{shannon1949communication,ahlswede1993common} that the cache memories do not increase the secrecy capacity. This paper considers an intermediate setting between these two extremes in which the adversary may have {\it{partial access}} to cache placement. 

The wiretap channel II (WTC-II) in \cite{WTCII_Wyner} provides a model for an adversary with partial access to the legitimate communication; in the form of a threshold on the time fraction during which the adversary is able to tap into the communication. Specifically, the model considers a noiseless legitimate channel and an adversary which {\it{selects}} a {\it{fixed-size}} subset of the transmitted symbols to noiselessly observe. \cite{WTCII_Wyner} has shown that, despite this ability to choose the locations of the tapped symbols, with proper coding, the adversary can be made no more powerful than nature, i.e., the secrecy capacity of the WTC-II is identical to that of a binary erasure wiretapper channel with the same fraction of erasures. \cite{nafea2015wiretap} has generalized the WTC-II to one with a discrete memoryless-- noisy-- legitimate channel, and derived inner and outer bounds for its capacity-equivocation region. The secrecy capacity for this model has been identified in \cite{goldfeld2015semantic}. In \cite{nafea2018new}, we have introduced a generalized wiretap model which includes both the classical wiretap \cite{WTCWyner} and wiretap II \cite{nafea2015wiretap} channels as special cases. This generalized model has been extended to multi-transmitter and multi-receiver networks in \cite{nafea2019generalizing,nafea2017new1,nafea2017new2}. In all these settings, the common theme is the robustness of stochastic wiretap encoding \cite{WTCWyner} against a type II adversary which can choose where to tap. 

In this paper, we introduce an adversary model of type II to a cache-aided communication setting. The adversary noiselessly observes a partial subset of its choice of the transmitted symbols over cache placement and/or delivery. We term this model the caching broadcast channel with a {\it{wire and cache tapping}} adversary of type II (CBC-WCT II). The legitimate terminals do not know whether cache placement, delivery, or both phases are tapped; the relative fractions of tapped symbols in each, nor their positions. Only the the overall size of the tapped set is known by the legitimate terminals. 

The challenge in caching stems from the fact that the transmitter, which has access to a library of files, has no knowledge about future demands of end users when designing their cache contents. This remains to be the case when security against an external adversary is concerned. Further, for the new model introduced in this paper, the adversary might tap into cache placement, delivery, or both, and where the tapping occurs is {\it{unknown}} to the legitimate terminals. We show that even under these unfavorable conditions, strong secrecy guarantees that are invariant to the positions of the tapped symbols varying between cache placement, delivery, or both phases, can be provided.

In coded caching literature up to date, the physical communication which populates the cache memories at end users is not considered in the problem formulation, due to the assumption of secure cache placement. For the setting we propose in this work, in order to model cache placement that is tapped by an adversary, we consider a length-$n$ communication block over a two-user broadcast WTC-II \cite{nafea2017new1}. The sizes of cache memories at the receivers are fixed in our setting. Introducing variable memory sizes for which a rate-memory tradeoff can be characterized, as in the usual setup for caching, requires considering additional communication blocks for cache placement. Being of future interest, we discuss this extension to multiple communication blocks for cache placement in Section \ref{Discussion}. We as well provide reasoning for our choice of the broadcast setting for cache placement. 
 
The main contributions of this paper are summarized as follows:
\begin{enumerate}
\item We introduce the notion of {\it{cache-tapping}} into the information theoretic models of coded caching, in which an adversary of type II is able to tap into a fixed-size subset of its choice of the symbols transmitted during either cache placement, delivery, or both phases. 
 
\item We characterize the strong secrecy capacity-- the maximum achievable file rate which keeps the overall library strongly secure-- for the instance of a transmitter's library with two files:
\begin{itemize}
\item We devise an achievability scheme which integrates wiretap coding \cite{WTCII_Wyner}, security embedding codes \cite{ly2012security,liang2014broadcast}, one-time pad keys \cite{shannon1949communication}, {\it{coded}} cache placement and {\it{uncoded}} delivery \cite{maddah2014fundamental}. 
\item We use a genie-aided upper bound in which the transmitter is provided with user demands before placement, rendering the model to a broadcast WTC-II \cite{nafea2017new1}, to establish the converse. 
\end{itemize}

\item We derive lower and upper bounds on the strong secrecy file rate when the library has more than two files: 
\begin{itemize}
\item We use the same channel coding scheme as for the two files case. However, the cache placement and delivery schemes we employ to achieve the rates are different. In particular, we utilize here {\it{uncoded}} cache placement and a {\it{partially coded}} delivery.  
\item We derive the upper bound in three steps: We (i) consider a transformed channel with an adversary which taps into a fraction of symbols equal to our model, but is only allowed to tap into the delivery phase. Since this adversary has a more restricted strategy space than the original one, its corresponding secrecy capacity is at least as large; (ii) use Sanov's theorem in method of types \cite[Theorem 11.4.1]{cover2006elements} to further upper bound the secrecy capacity of the restricted adversary model by the secrecy capacity when the adversary encounters a discrete memoryless binary erasure channel, and finally (iii) upper bound the secrecy capacity of the discrete memoryless model by that of a single receiver setting in which the receiver requests two files from the library.
\end{itemize}
\end{enumerate}

The remainder of the paper is organized as follows. Section \ref{ChannelModel} describes the communication system proposed in this paper. Section \ref{MainResult} presents the main results. The proofs of these results are provided in Sections \ref{Proof_Thm1}, \ref{Proof_Thm2}, and \ref{Proof_Thm3}. Section \ref{Discussion} provides a discussion about the communication model in question, the presented results, and the extension of our model to arbitrary number of users and to variable memory sizes. Section \ref{Con} concludes the paper. 

\begin{figure}
\centering
\includegraphics[scale=0.62]{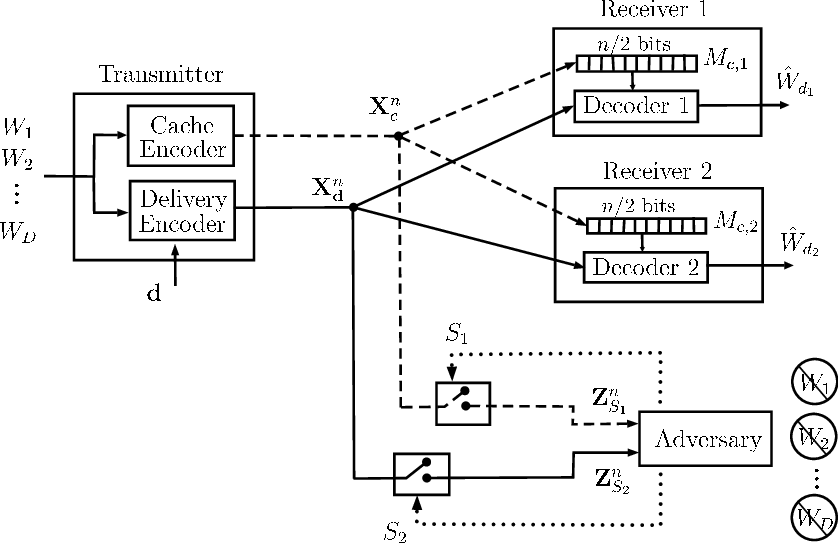}
\caption{The caching broadcast channel with a wire and cache tapping adversary of type II (CBC-WCT II). The adversary chooses tapping sets $S_1$ and $S_2$ in placement and delivery.}
\label{fig:sysmodel_1}
\end{figure}

\section{System Model}{\label{ChannelModel}}
We remark the notation we use throughout the paper. $\mathbb{N}$, $\mathbb{Z}$, $\mathbb{R}$ denote the sets of natural, integer, real numbers, respectively. For $a,b\in\mathbb{R}$, $[a:b]$ denotes the set of integers $\{i\in\mathbb{N}: a\leq i\leq b\}$. $A_{[1:n]}$ denotes the sequence of variables $\{A_1,A_2,\cdots,A_n\}$. For two sets $\setA_1$, $\setA_2$; $\setA_1\times \setA_2$ denotes their Cartesian product. $\setA^T$ denotes the $T$-fold Cartesian product of the set $\setA$. For $W_1,W_2\in[1:M]$, $W_1\oplus W_2$ denotes the bit-wise XOR on the binary strings corresponding to $W_1$, $W_2$. $\mathbbm{1}_{\setA}$ denotes the indicator function for the event $\setA$. $\mathbb{D}(p_x||q_x)$ denotes the Kullback-Leibler divergence between the probability distributions $p_x$, $q_x$, defined on the same probability space. $\{\epsilon_n\}_{n\geq 1}$ denotes a sequence of positive real numbers s.t. $\epsilon_n \rightarrow 0$ as $n \rightarrow\infty$. 

Consider the communication system depicted in Fig. \ref{fig:sysmodel_1}, in which the adversary is able to tap into both the cache placement and delivery transmissions. The transmitter observes $D\geq 2$ independent messages (files), $W_1,W_2,\cdots,W_D$, each of which is uniformly distributed over $[1:2^{nR_s}]$. Each receiver has a cache memory of size $\frac{n}{2}$ bits. The communication occurs over two phases: cache placement and delivery. The broadcast channel is noiseless during both phases. The communication model is described as follows:

{\it{Cache placement phase:}}
The transmitter broadcasts a length-$n$ binary signal, $\bX_c^n\in\{0,1\}^n$, to both receivers. The codeword $\bX_c^n$ is a function of the library files; $\bX_c^n\triangleq f_c(W_{[1:D]})$. The transmitter does not know the receiver demands during cache placement \cite{maddah2014fundamental}. Each receiver has a cache memory of size $\frac{n}{2}$ bits in which they store a function of $\bX_c^n$, $M_{c,j}\triangleq f_{c,j}(\bX_c^n)$; $f_{c,j}: \{0,1\}^n\mapsto [1:2^{\frac{n}{2}}]$, $j=1,2$. 

{\it{Delivery phase:}}
At the beginning of this phase, the two receivers announce their demands $\bd\triangleq (d_1,d_2)\in [1:D]^2$ to the transmitter. To satisfy these demands, the transmitter encodes $W_{[1:D]}$ and $\bd$ into a binary codeword $\bX_{\bd}^n\in\{0,1\}^n$: For each $\bd\in [1:D]^2$, the transmitter uses the encoder $f_{\bd}: [1:2^{nR_s}]^D \mapsto\{0,1\}^n$ and sends the codeword $\bX_{\bd}^n\triangleq f_{\bd}(W_{[1:D]})$.

{\it{Decoding:}} Receiver $j$ uses the decoder $g_{\bd,j}:[1:2^{\frac{n}{2}}]\times \{0,1\}^n \mapsto [1:2^{nR_s}]$ to output the estimate $\hat{W}_{d_j}\triangleq g_{\bd,j}(f_{c,j}(\bX_c^n),\bX_{\bd}^n)$ of its desired message $W_{d_j}$; $j=1,2$. 

{\it{Adversary model:}} The adversary chooses two subsets $S_1,S_2\subseteq [1:n]$. The size of the sum of cardinalities of $S_1$ and $S_2$ is fixed: For $|S_1|=\mu_1$, $|S_2|=\mu_2$, $\mu_1,\mu_2\leq n$, we have $\mu_1+\mu_2= \mu$. The subsets $S_1,S_2$ indicate the positions tapped by the adversary during cache placement and delivery, respectively. Over the two phases, the adversary observes the length-$2n$ sequence $\bZ_{S}^{2n}= [\bZ_{S_1}^n,\bZ_{S_2}^n]\in\setZ^{2n}$, where $\bZ_{S_j}^n\triangleq [Z_{S_j,1},\cdots,Z_{S_j,n}]\in\setZ^{n}$, $j=1,2$,
\begin{align}
\label{eq:Z_S_1_2}
&Z_{S_1,i}=\begin{cases}
X_{c,i},\; i\in S_1\\
?,\; i\notin S_1
\end{cases}, \;\;
Z_{S_2,i}=\begin{cases}
X_{\bd,i},\; i\in S_2\\
?,\; i\notin S_2.
\end{cases}
\end{align}
The alphabet is $\setZ=\{0,1,?\}$, where $``?"$ denotes an erasure. 

The legitimate terminals know neither the realizations of $S_1,S_2$, nor the values of $\mu_1,\mu_2$. Only $\mu$ is known. Let $\alpha_1=\frac{\mu_1}{n}$, $\alpha_2=\frac{\mu_2}{n}$, be the fractions of the tapped symbols in the cache placement and delivery, and let $\alpha=\alpha_1+\alpha_2$ be the overall tapped ratio. Note that $\alpha_1,\alpha_2\in [0,1]$ and $\alpha\in (0,2]$.

\begin{remark}
We consider that $\alpha>0$, i.e., the adversary is present. For $\alpha=0$, i.e., no adversary, the problem considered in this paper has been extensively studied in the literature, see for example \cite{maddah2014fundamental,tian2018symmetry,ibrahim2019coded,cao2019coded}.
\end{remark}

A channel code $\setC_{2n}$ for this model consists of 
\begin{itemize}
\item $D$ message sets; $\setW_{l}\triangleq[1:2^{nR_s}]$, $l=1,2,\cdots,D$,
\item Cache encoder; $f_c:[1:2^{nR_s}]^D\mapsto \{0,1\}^n$,
\item Cache decoders; $f_{c,j}:\{0,1\}^n\mapsto [1:2^{\frac{n}{2}}]$, $j=1,2$,
\item Delivery encoders;  $f_{\bd}:[1:2^{nR_s}]^D\mapsto \{0,1\}^n$; where $\bd \in [1:D]^2$,
\item Decoders; $g_{\bd,j}:[1:2^{\frac{n}{2}}]\times\{0,1\}^n\mapsto [1:2^{nR_s}]$; where $j=1,2,\; \bd\in [1:D]^2$.
\end{itemize}
The file rate $R_s$ is {\it{achievable with strong secrecy}} if there is a sequence of codes $\{\setC_{2n}\}_{n\geq 1}$ such that
\begin{align}
\label{eq:reiaility}
&\limitn \max_{\bd\in [1:D]^2}\mathbb{P}\Big(\bigcup_{j=1,2}(\hat{W}_{d_j}\neq W_{d_j})\Big)=0 \;\;\; \textbf{(Reliability)},\\
\label{eq:secrecy}
&\limitn\; \underset{\begin{subarray}{c} S_1,S_2\subseteq [1:n]:\\|S_1|+|S_2|\leq \mu\end{subarray}}\max I(W_{[1:D]};\bZ_{S_1}^n,\bZ_{S_2}^n)=0\; \textbf{(Strong Secrecy)}.
\end{align}
That is, $R_s$ is the {\it{symmetric secure}} file rate, under any demand vector and adversarial strategy. The strong secrecy capacity $C_s$ is the the supremum of all achievable $R_s$. 

\begin{remark}
While we consider the file rate $R_s$ which guarantees reliability for the worst-case demand vector, the average rate for which there exists a prior distribution on the demands has been studied in coded caching literature  as well; see for example \cite{niesen2017coded,lim2017information,ji2017order}.
\end{remark}

\begin{remark}
The condition in (\ref{eq:secrecy}) guarantees strong secrecy against all possible strategies for the adversary, i.e., choices of $S_1,S_2$ that satisfy the condition $|S_1|+|S_2|\leq \mu$.
\end{remark}

\section{Main Results}\label{MainResult}
For clarity of exposition, we first study the model in Section \ref{ChannelModel} when the transmitter's library has two files; $D=2$. We then extend the ideas and analysis to $D>2$. For $D>2$, we use a similar channel coding scheme to that we construct for $D=2$, but the placement and delivery schemes that achieve the best rates are different. The following theorem presents the strong secrecy capacity for $D=2$.
\begin{theorem}\label{thm:Thm1}
For $0<\alpha\leq 2$ and $D=2$, the strong secrecy capacity for the caching broadcast channel with a wire and cache tapping adversary of type II (CBC-WCT II), described in Section \ref{ChannelModel}, is given by
\begin{align}
\label{eq:sum_secrecy_capacity_D=2}
C_s(\alpha)=1-\frac{\alpha}{2}.
\end{align}
\end{theorem}

\begin{Proof}
The proof is provided in Section \ref{Proof_Thm1}.
\end{Proof}

Theorem \ref{thm:Thm2} below presents an achievable strong secrecy file rate for $D>2$. 
\begin{theorem}\label{thm:Thm2}
For $0<\alpha\leq 2$, $D>2$, an achievable strong secrecy file rate for the CBC-WCT II is 
\begin{align}
\label{eq:sum_secrecy_rate_D>2}
R_s(\alpha)\geq \begin{cases}
\frac{1}{2}+\frac{3(1-\alpha)}{4D},\qquad 0<\alpha<1 \\
1-\frac{\alpha}{2},\qquad 1\leq \alpha\leq 2.\end{cases}
\end{align}
\end{theorem}
\begin{Proof}
The proof is provided in Section \ref{Proof_Thm2}.
\end{Proof}

The following theorem upper bounds the secure file rate when $D>2$.  
\begin{theorem}\label{thm:Thm3}
For $0<\alpha\leq 2$, $D>2$, the strong secrecy file rate for the CBC-WCT II is upper bounded as
\begin{align}
\label{eq:sum_secrecy_bound_D>2}
R_s(\alpha)\leq \begin{cases}
\frac{1}{2}+\frac{2D-1}{2D(D-1)}(1-\alpha),\qquad 0< \alpha<1\\
1-\frac{\alpha}{2},\qquad 1\leq \alpha\leq 2.\end{cases}
\end{align}
\end{theorem}
\begin{Proof}
The proof is provided in Section \ref{Proof_Thm3}.
\end{Proof}

\begin{figure}
\centering
\includegraphics[scale=0.34]{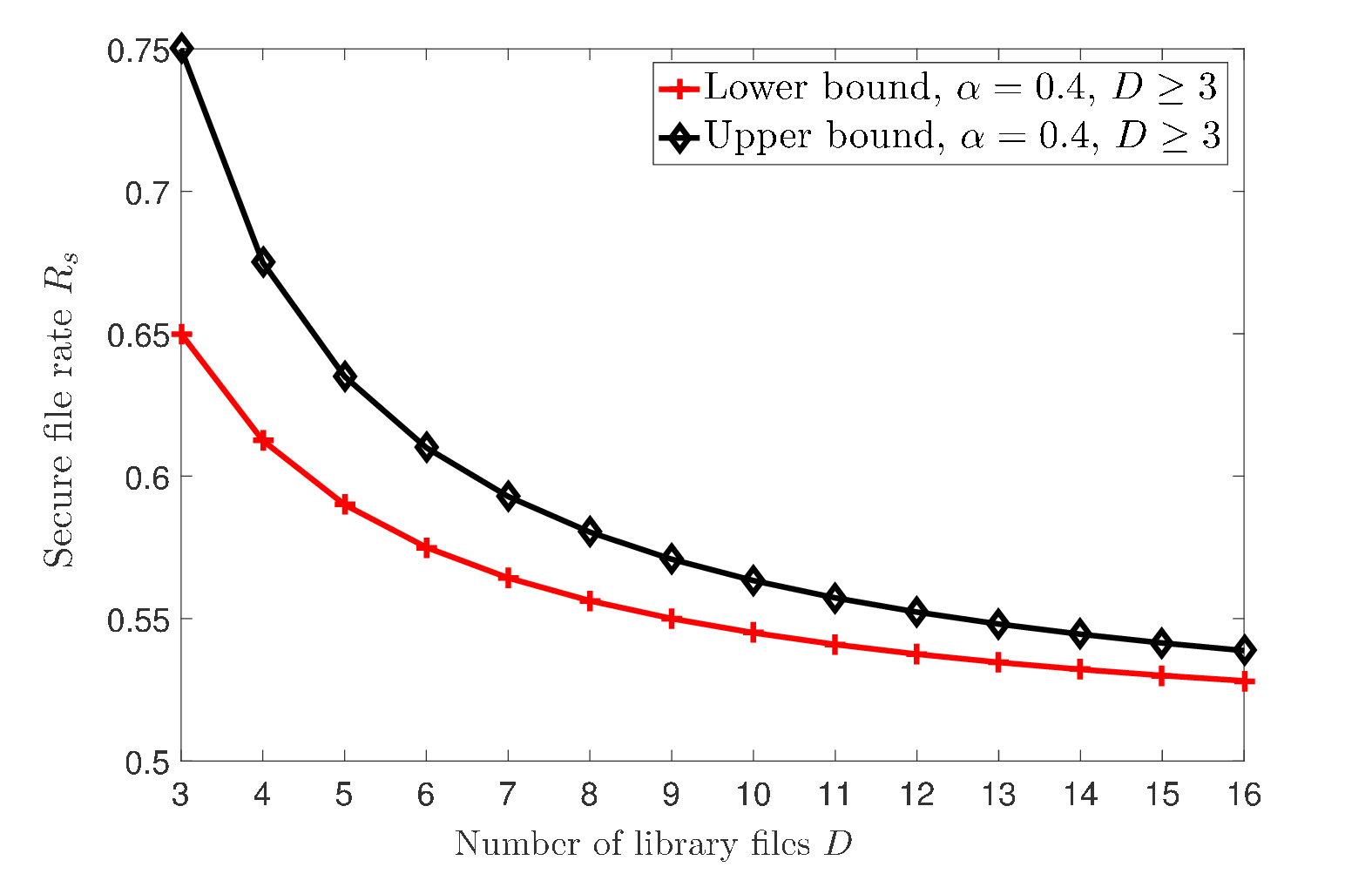}
\caption{Bounds on the achievable strong secrecy file rate $R_s$, when $\alpha=0.4$ and $D\geq 3$.}
\label{fig:lower_upper_bounds}
\end{figure}

The following corollary is immediate from Theorems \ref{thm:Thm1}--\ref{thm:Thm3}. 
\begin{corollary}\label{Cor:Cor1}
For $1\leq\alpha\leq 2$, that is when the adversary can tap longer than one phase of communication, the strong secrecy capacity for the CBC-WCT II is 
\begin{align}
\label{eq:sum_secrecy_capacity_alpha>1}
C_s(\alpha)=1-\frac{\alpha}{2}.
\end{align}
\end{corollary}

\begin{remark}
When $\alpha\in [1,2]$ ($n\leq \mu\leq 2n$), two possible strategies for the adversary are $\{S_1=[1:n], S_2\subset [1:n]\}$ and $\{S_1\subset [1:n], S_2=[1:n]\}$; the adversary taps into all symbols in one phase and a subset of symbols in the other. Interestingly, the secrecy capacity for this range of $\alpha$ is $1-\frac{\alpha}{2}$ for any library size. As we shall see in Sections \ref{Achievability_2} and \ref{Proof_Thm2}, such an adversary limits the communication for cache placement to exchanging additional randomness  (key bits) that allows for communicating a positive secure rate over the two phases. The cache memories are thus not used to store any information, and hence the lack of knowledge about user demands during cache placement is immaterial.
\end{remark}

Unlike for $1\leq \alpha\leq 2$, for $0<\alpha<1$, the lower and upper bounds in (\ref{eq:sum_secrecy_rate_D>2}) and (\ref{eq:sum_secrecy_bound_D>2}) have a gap. For illustration purposes, these bounds are plotted for $\alpha=0.4$ in Fig. \ref{fig:lower_upper_bounds}.

\begin{remark}
When $\alpha=0$ (no adversary), our achievability scheme for $D>2$ in Section \ref{Proof_Thm2} reduces to the achievability scheme in \cite{maddah2014fundamental}, which is shown to achieve the optimal rate-memory tradeoff for this case \cite{tian2018symmetry,cao2019coded}. However, the upper bound for $D>2$ derived in this work is to address the intricacies of the adversarial model and is useful only when the adversary is present ($\alpha>0$); (\ref{eq:sum_secrecy_bound_D>2}) is loose when $\alpha=0$.
\end{remark}

\section{Proof of Theorem \ref{thm:Thm1}}\label{Proof_Thm1}
In this section, we prove Theorem \ref{thm:Thm1} which identifies the strong secrecy capacity when $D=2$. Recall that the demand vector is denoted by $\bd=(d_1,d_2)$, where $d_1,d_2\in\{1,2\}$.

\subsection{Converse}\label{Converse}
When $\bd$ is known to the transmitter during cache placement, the model in Theorem \ref{thm:Thm1} reduces to a broadcast wiretap channel II (WTC-II), over a length-$2n$ communication block. The strong sum secrecy rate for that model, $2R_s$, is upper bounded by
\begin{align}
\label{eq:known_demand_tapped_caching}
2R_s\leq 2-\alpha,
\end{align}
which follows from our recent work \cite[Theorem 1]{nafea2017new1}. Note that (\ref{eq:known_demand_tapped_caching}) holds for any $\bd=(d_1,d_2)$ such that $d_1\neq d_2$, which represents the worst-case demands. Since $\bd$ is unknown for the model in consideration, $1-\frac{\alpha}{2}$ is an upper bound for its strong secrecy capacity. 

\subsection{Restricted Adversary Models as Building Blocks}\label{Restricted_Models}
Before proceeding with the achievability proof, it is relevant to take a step back and investigate the secrecy capacity when a known fraction of cache placement, a known fraction of delivery, or both, is tapped. In particular, we consider that the adversary taps into (i) cache placement only, (ii) delivery only, or (iii) both and the relative fractions of tapped symbols in each are known. For these three models, we show that the strong secrecy file rate in (\ref{eq:sum_secrecy_capacity_D=2}), i.e., $1-\frac{\alpha}{2}$, is achievable, and hence determines their strong secrecy capacities. We then use these models as building blocks for when the relative fractions are {\it{unknown}}, and provide the achievability proof in Sections \ref{Achievability_1} and \ref{Achievability_2}.

\subsubsection{Setting $1$: The adversary taps into cache placement only} This setting corresponds to $\alpha_1=\alpha$ ($\alpha_2=0$) and $|S_1|=\mu$ ($S_2=\varnothing$). The transmitter and receivers know that $\alpha_1=\alpha$. We show that $1-\frac{\alpha}{2}$ is an achievable strong secrecy file rate for this setting. 

The transmitter divides $W_l,\;l=1,2,$ into three independent messages, $W_l^{(1)}$, $W_l^{(2)}$, $W_{l,s}$; $W_l^{(1)}$, $W_l^{(2)}$ are uniform over $[1:2^{n\frac{1-\alpha-\epsilon_n}{2}}]$, $W_{l,s}$ is uniform over $[1:2^{n\frac{\alpha+\epsilon_n}{2}}]$. Define
\begin{align}
\label{eq:example1_1}
\nonumber &M_c\triangleq\{M_{c,1},M_{c,2}\};\\
&M_{c,1}= W_1^{(1)}\oplus W_2^{(1)},\quad M_{c,2}= W_1^{(2)}\oplus W_2^{(2)},\\
\label{eq:example1_2}
&M_{\bd}\triangleq\{W_{d_1}^{(2)},W_{d_2}^{(1)},W_{d_1,s},W_{d_2,s}\},
\end{align}
where $M_c$ and $M_{\bd}$ are the messages sent by the transmitter during cache placement and delivery phases, respectively. Specifically, during placement, the transmitter maps $M_c$ into $\bX_c^n$ using stochastic wiretap coding \cite{WTCWyner}. Since the rate of $M_c$ is less than $1-\alpha$, $M_c$ is strongly secure against the adversary that observes $n\alpha$ symbols of $\bX_c^n$ \cite{nafea2018new,goldfeld2015semantic}. During delivery, the transmitter sends $\bX_{\bd}^{n}$ as the binary representation of $M_{\bd}$ which is of length $n$ bits, since the delivery phase is noiseless and secure. 

Using $\bX_c^n$, noiselessly received during placement, receiver $j$, $j=1,2,$ recovers $M_{c,j}$ and stores it in its cache memory. The size of $M_{c,j}$ is smaller than $\frac{n}{2}$ bits, which is the cache size at each receiver. Using $\bX_{\bd}^n$, received noiselessly during delivery, both receivers recover $M_{\bd}$. Using $M_{\bd}$, along with its cache contents, $M_{c,j}$, and for $n$ sufficiently large\footnote{Large block-length $n$ is needed to ensure a valid subpacketization of the file $W_l$ into the sub-files $\{W_l^{(1)},W_l^{(2)},W_{l,s}\}$, for $l=1,2$. That is, a bijective map between the file and its sub-files is preserved.}, receiver $j$ correctly recovers its desired message $W_{d_j}$, $j=1,2$. 

For secrecy, we show in Appendix \ref{AppendixA} that (\ref{eq:secrecy}) is satisfied. Since $\epsilon_n \rightarrow 0$ as $n \rightarrow\infty$, the achievable strong secrecy file rate is given by 
\begin{align}
\label{eq:example1_4}
R_s(\alpha)=2\times \frac{1-\alpha}{2}+\frac{\alpha}{2}=1-\frac{\alpha}{2}.
\end{align}

\subsubsection{Setting $2$: The adversary taps into the delivery only}
This setting corresponds to $\alpha_1=0$ and $\alpha_2=\alpha$, and the transmitter and receivers possess this knowledge. Once again, we show that $1-\frac{\alpha}{2}$ is an achievable strong secrecy file rate. The transmitter (i) performs the same division of $W_l,\;l=1,2,$ as in Setting $1$, (ii) generates the keys $K_1$, $K_2$, each is uniform over $[1:2^{n\frac{\alpha+\epsilon_n}{2}}]$, independent from one another and from $W_1$, $W_2$. Define $M_c$, $M_{\bd}$, $\tilde{M}_{\bd}$, as follows:
\begin{align}
\label{eq:example2_2}
\nonumber&M_c=\{M_{c,1},M_{c,2}\};\;\;M_{c,1}=\{W_1^{(1)}\oplus W_2^{(1)},K_1\},\\
&M_{c,2}=\{W_1^{(2)}\oplus W_2^{(2)},K_2\},\\
\label{eq:example2_3}
&M_{\bd}=\{W_{d_1}^{(2)},W_{d_2}^{(1)}\},\; \tilde{M}_{\bd}=\{W_{d_1,s}\oplus K_1,W_{d_2,s}\oplus K_2\}.
\end{align}

During placement, the transmitter sends $\bX_c^n$ as the binary representation of $M_c$, and receiver $j$, $j=1,2,$ stores $M_{c,j}$ in its cache memory. During delivery, the transmitter encodes $M_{\bd}$ into $\bX_{\bd}^n$ using wiretap coding, while using $\tilde{M}_{\bd}$ as the randomization message. Receiver $j$ recovers $M_{\bd}$, $\tilde{M}_{\bd}$, using which, along with $M_{c,j}$, it correctly decodes $W_{d_j}$, for sufficiently large $n$. By contrast, the adversary can only obtain $\tilde{M}_{\bd}$ using which it can gain no information about $W_1$, $W_2$. In Appendix \ref{AppendixB}, we show that (\ref{eq:secrecy}) is satisfied. The achievable strong secrecy file rate is again $1-\frac{\alpha}{2}$.

\subsubsection{Setting $3$: The legitimate terminals know the values of $\alpha_1$ and $\alpha_2$}
For this setting, neither $\alpha_1=0$ nor $\alpha_2=0$. However, the transmitter and receivers know the values of $\alpha_1$, $\alpha_2$. Under these assumptions, the scheme which achieves the secrecy rate $1-\frac{\alpha}{2}$ depends on whether $\alpha_1\geq \alpha_2$. For $\alpha_1\geq \alpha_2$ ($\alpha_1<\alpha_2$), we use an achievability scheme similar to Setting $1$ (Setting $2$). 

{\it{Case $1$: $\alpha_1\geq \alpha_2$:}} The transmitter divides $W_l,\;l=1,2,$ into the independent messages $\{W_l^{(1)},W_l^{(2)},W_{l,s}\}$; $W_l^{(1)},W_l^{(2)}$ are uniform over $[1:2^{n\frac{1-\alpha_1-\epsilon_n}{2}}]$ and $W_{l,s}$ is uniform over $[1:2^{n\frac{{\alpha_1-\alpha_2}}{2}}]$. The transmitter forms $M_c$, $M_{\bd}$, as in (\ref{eq:example1_1}), (\ref{eq:example1_2}), and uses wiretap coding to map them into $\bX_c^n$, $\bX_{\bd}^n$, respectively. As in setting $1$, receiver $j$ recovers $W_{d_j}$. For the secrecy constraint, note that $M_c$, $M_{\bd}$ are independent, and their rates are $1-\alpha_1-\epsilon_n$, $1-\alpha_2-\epsilon_n$, respectively. Thus, wiretap coding strongly secures both $M_c$ and $M_{\bd}$ against the adversary. We show in Appendix \ref{AppendixC} that (\ref{eq:secrecy}) is satisfied. The achievable strong secrecy file rate is 
$R_s(\alpha)=2\times \left(\frac{1-\alpha_1}{2}\right)+\frac{\alpha_1-\alpha_2}{2}=1-\frac{\alpha}{2}$.

{\it{Case $2$: $\alpha_1<\alpha_2$:}} The transmitter (i) divides $W_l$ into $\{W_l^{(1)},W_l^{(2)},W_{l,s}\}$, where $W_l^{(1)},W_l^{(2)}$ are uniform over $[1:2^{n\frac{1-\alpha_2-\epsilon_n}{2}}]$ and $W_{l,s}$ is uniform over $[1:2^{n\frac{{\alpha_2-\alpha_1}}{2}}]$, $l=1,2$; (ii) generates the keys $K_1,K_2$, uniformly over $[1:2^{n\frac{\alpha_2-\alpha_1}{2}}]$ and independently from $W_1,W_2$; (iii) forms $M_c$ as in (\ref{eq:example2_2}) and encodes it into $\bX_c^n$ using wiretap coding; (iv) forms $M_{\bd}$ as in (\ref{eq:example2_3}) and forms $\tilde{M}_{\bd}$ as 
\begin{align}
&\tilde{M}_{\bd}=\{W_{d_1,s}\oplus K_1,W_{d_2,s}\oplus K_2,\tilde{W}\},
\end{align}
$\tilde{W}$ is independent from all other variables and uniform over $[1:2^{n(\alpha_1+\epsilon_n)}]$, and finally (v) encodes $M_{\bd}$ into $\bX_{\bd}^n$ using wiretap coding, while using $\tilde{M}_{\bd}$ as the randomization message. 

As in Setting $2$, receiver $j$ correctly recovers $W_{d_j}$, and the adversary can only recover $\tilde{M}_{\bd}$ using which it gains no information about $W_1$ and $W_2$. In Appendix \ref{AppendixD}, we show (\ref{eq:secrecy}) is satisfied. The achievable secrecy rate is 
$R_s(\alpha)=2\times \left(\frac{1-\alpha_2}{2}\right)+\frac{\alpha_2-\alpha_1}{2}=1-\frac{\alpha}{2}$. 

With the aforementioned settings, we showed that the same secrecy rate, $1-\frac{\alpha}{2}$, is achievable irrespective of where the adversary taps as long as $\alpha_1$ and $\alpha_2$ are known. The question then arises whether the lack of knowledge about relative fractions of tapped symbols would decrease the secrecy capacity. The following setting we propose provides a hint on the answer.

\subsubsection{Setting $4$: Either $\alpha_1=0$ or $\alpha_2=0$, the legitimate terminals do not know which is zero}
The adversary taps into either cache placement or delivery, but not both. The legitimate terminals {\it{do not}} know which phase is tapped. We show that the strong secrecy rate $1-\frac{\alpha}{2}$ is again achievable. 

The transmitter performs the same division of $W_1$ and $W_2$ as in Settings $1$, $2$, and generates independent keys $K_1$ and $K_2$ as in Setting $2$. Define 
\begin{align}
\label{eq:M_c}
\nonumber &M_c=\{M_{c,1},M_{c,2}\};\\ &M_{c,1}=W_1^{(1)}\oplus W_2^{(1)},\;\;\; M_{c,2}=W_1^{(2)}\oplus W_2^{(2)},\\
\label{eq:M_c_tilde}
&\tilde{M}_c=\{\tilde{M}_{c,1},\tilde{M}_{c,2}\};\;\;
\tilde{M}_{c,1}=K_1,\; \tilde{M}_{c,2}=K_2\\
\label{eq:M_d}
&M_{\bd}=\{W_{d_1}^{(2)},W_{d_2}^{(1)}\},\\
\label{eq:M_d_tilde}
\nonumber &\tilde{M}_{\bd}=\{\tilde{M}_{\bd,1},\tilde{M}_{\bd,2}\};\; \tilde{M}_{\bd,1}=W_{d_1,s}\oplus K_1,\\
& \tilde{M}_{\bd,2}=W_{d_2,s}\oplus K_2.
\end{align}

During placement, the transmitter encodes $M_c$ into $\bX_c^n$ using wiretap coding, while using $\tilde{M}_c$ as the randomization message. Receiver $j$, $j=1,2,$ stores $M_{c,j}$, $\tilde{M}_{c,j}$, in its cache. During delivery, the transmitter uses wiretap coding to encode $M_{\bd}$ into $\bX_{\bd}^n$, while using $\tilde{M}_{\bd}$ as the randomization message. Using its cache contents, and $M_{\bd}$, $\tilde{M}_{\bd}$, receiver $j$ correctly decodes $W_{d_j}$. By contrast, the adversary can only recover either $\{K_1,K_2\}$ or $\{W_{d_1,s}\oplus K_1,W_{d_2,s}\oplus K_2\}$, but not both, using which it gains no information about $W_1$, $W_2$. We show in Appendix \ref{AppendixE} that (\ref{eq:secrecy}) is satisfied. The achievable strong secrecy rate is $1-\frac{\alpha}{2}$. 

The lack of knowledge about which phase is tapped is countered by encrypting pieces of information, $\{W_{d_1,s},W_{d_2,s}\}$, with one-time pad keys, $K_1$, $K_2$, while ensuring the adversary only recovers either the keys or the encrypted bits but not both; using which it gains no information about $W_1$, $W_2$. 

We next generalize this idea to tackle the case when the adversary taps into both phases, with no knowledge about the relative fractions of tapped symbols in each (the model in Fig. \ref{fig:sysmodel_1}). Before continuing, let us first describe {\it{security embedding codes}} \cite{liang2014broadcast,ly2012security}. The transmitted message is split into a number of layers, corresponding to different security levels. All layers of the message are encoded into a single codeword in an embedded fashion; each layer corresponds to one index identifying the codeword. Lower security-level layers serve as randomization (stochastic coding) for protecting higher security-level layers. The layers that can be securely transmitted are determined by the wiretapper's channel state. 

Similar to \cite{liang2014broadcast} where the uncertainty about the wiretapper's channel is treated using security embedding codes, here, in each phase, we construct an embedding code in which $n\alpha$ single-bit layers are embedded into one another. Doing so, we ensure that, no matter what the values for $\alpha_1,\alpha_2$ are, the adversary gets no more than $n\alpha_1$ bits from cache placement, and $n\alpha_2$ bits from delivery. By designing what the adversary recovers to be either a set of key bits and/or information bits encrypted with a distinct set of key bits, we guarantee no information on the messages is asymptotically leaked to the adversary. We thus prove that the lack of knowledge about relative fractions of tapped symbols {\it{does not decrease}} the secrecy capacity. 

\subsection{Achievability for $\alpha\in(0,1)$:}\label{Achievability_1}
We are now ready to present the achievability for the general model when $D=2$. Consider first $\alpha\in(0,1)$. For simplicity, assume that $\frac{n\alpha_1}{2}=\frac{\mu_1}{2}$ and $\frac{n\alpha_2}{2}=\frac{\mu_2}{2}$ are integers. A minor modification to the analysis can be adopted otherwise. The transmitter (i) divides $W_l$, $l=1,2,$ into the independent messages $W_l^{(1)}$, $W_l^{(2)}$, $W_{l,s}$; $W_l^{(1)}$, $W_l^{(2)}$ are uniform over $[1:2^{n\frac{1-\alpha}{2}}]$, and $W_{l,s}$ is uniform over $[1:2^{n\frac{\alpha}{2}}]$; (ii) generates the  independent keys $K_1,K_2$, uniform over $[1:2^{n\frac{\alpha}{2}}]$, and independent from $W_1,W_2$. For simplicity of exposition, we have ignored the small rate reduction $\epsilon_n$ at this stage, as we will introduce this later into the security analysis. The main ideas of the achievability proof are:

\begin{enumerate}
\item The transmitter uses wiretap coding with a randomization message of size $n\alpha$ bits in {\it{both placement and delivery}}. As the adversary does not tap into more than $n\alpha$ bits in each phase, a secure transmission rate of $1-\alpha$ is achievable, as long as the randomization messages in the two phases are independent. Using coded placement for $\{W_l^{(1)},W_l^{(2)}\}_{l=1,2}$, a secure file rate of $1-\alpha$ can be achieved.

\item The randomization messages over the two phases can deliver additional secure information, of rate $\frac{\alpha}{2}$ per file, via encryption. The overall achievable file rate is $R_s=1-\frac{\alpha}{2}$. We use $K_1$, $K_2$, as the randomization message for placement. Along with wiretap coding, we employ a security embedding code \cite{ly2012security}, using bits of $K_1$, $K_2$, in a manner that allows the adversary to recover only the last $n\frac{\alpha_1}{2}$ bits from each. During delivery, we encrypt $W_{d_1,s}$, $W_{d_2,s}$, with $K_1$, $K_2$, and use this encrypted information as the randomization message. We employ again a security embedding code in the {\it{reverse order}} so that the adversary recovers only the first $n\frac{\alpha_2}{2}$ bits from each of $W_{d_1,s}\oplus K_1$, $W_{d_2,s}\oplus K_2$.

\item With the aforementioned construction, the adversary, for any values of $\alpha_1$, $\alpha_2$ it chooses, can only recover a set of key bits and/or a set of information bits encrypted with other key bits. Due to the {\it{reversed embedding order}}, the adversary does not obtain, during delivery, any message bits encrypted with key bits it has seen during placement. In addition, since $\{K_1,K_2\}$ is independent from $\{W_{d_1,s}\oplus K_1,W_{d_2,s}\oplus K_2\}$, and is an independent sequence, the adversary can not use the revealed key bits during cache placement to obtain any information about the bits of $W_{d_1,s}\oplus K_1$, $W_{d_2,s}\oplus K_2$ that need to be securely transmitted during delivery. 
\end{enumerate} 

\begin{figure*}
\centering
\includegraphics[scale=0.63]{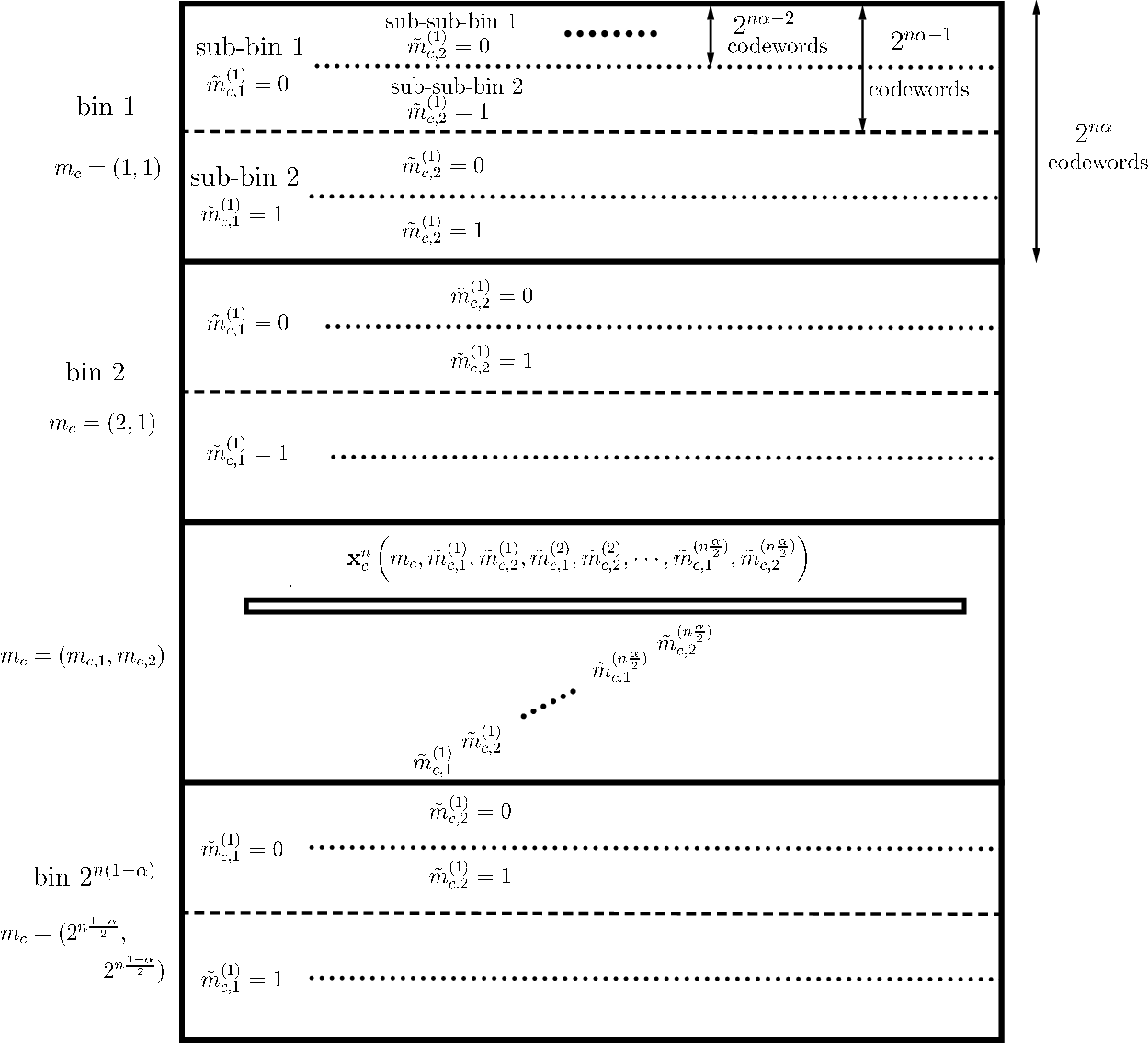}
\caption{Codebook construction for the cache placement phase, $\setC_{c,n}$.}
\label{fig:code_construction_1}
\end{figure*}

We now explain the achievability scheme in more detail. Let $M_c$, $\tilde{M}_c$ be as in (\ref{eq:M_c}), (\ref{eq:M_c_tilde}). $M_c$ represents the message to be securely transmitted during placement, regardless of the adversary's choice of $\alpha_1$. $\tilde{M}_c$ represents the randomization message used for wiretap coding. The transmitter further divides $\tilde{M}_{c,1}$, $\tilde{M}_{c,2}$ into sequences of independent bits, $\{\tilde{M}_{c,1}^{(1)},\cdots,\tilde{M}_{c,1}^{(n\frac{\alpha}{2})}\}$, $\{\tilde{M}_{c,2}^{(1)},\cdots,\tilde{M}_{c,2}^{(n\frac{\alpha}{2})}\}$, and generates $\bX_c^n$ as: 

{\it{Cache Placement Codebook Generation:}} Let $m_c$, $\tilde{m}_{c,1}=\{\tilde{m}_{c,1}^{(1)},\cdots,\tilde{m}_{c,1}^{(n\frac{\alpha}{2})}\}$, $\tilde{m}_{c,2}=\{\tilde{m}_{c,2}^{(1)},\cdots,\tilde{m}_{c,2}^{(n\frac{\alpha}{2})}\}$ be the realizations of $M_c$, $\tilde{M}_{c,1}$, $\tilde{M}_{c,2}$. We construct the codebook $\setC_{c,n}$, from which $\bX_c^n$ is drawn, as follows. We randomly and independently distribute all the possible $2^{n}$ length-$n$ binary sequences into $2^{n(1-\alpha)}$ bins, indexed by $m_c \in [1:2^{n\frac{1-\alpha}{2}}]^2$, and each contains $2^{n\alpha}$ codewords. Further, we randomly and independently divide each bin $m_c$ into two sub-bins, indexed by $\tilde{m}_{c,1}^{(1)}$, and each contains $2^{n\alpha-1}$ codewords. The two sub-bins $\tilde{m}_{c,1}^{(1)}$ are further divided into smaller bins, indexed by $\tilde{m}_{c,2}^{(1)}$, and each contains $2^{n\alpha-2}$ codewords. The process continues, going over $\tilde{m}_{c,1}^{(2)}$, $\tilde{m}_{c,2}^{(2)}$, $\cdots$, $\tilde{m}_{c,1}^{(n\frac{\alpha}{2}-1)}$, $\tilde{m}_{c,2}^{(n\frac{\alpha}{2}-1)}$, $\tilde{m}_{c,1}^{(n\frac{\alpha}{2})}$, until the remaining two codewords, after each sequence of divisions, are indexed by $\tilde{m}_{c,2}^{(n\frac{\alpha}{2})}$. $\setC_{c,n}$ is described in Fig. \ref{fig:code_construction_1}. 

\begin{remark}
An alternative representation of the former binning procedure is: Each of the $2^{n\alpha}$ codewords in the bin $m_c$, $m_c \in [1:2^{n\frac{1-\alpha}{2}}]^2$, is randomly assigned to an index $\{\tilde{m}_{c,1}^{(1)},\tilde{m}_{c,2}^{(1)},\cdots,\tilde{m}_{c,1}^{(n\frac{\alpha}{2})},\tilde{m}_{c,2}^{(n\frac{\alpha}{2})}\}$. We chose to present the former description to provide a more detailed explanation of the embedding structure; in particular, the order of embedding which is a critical component in the achievability scheme.
\end{remark}

{\it{Cache Encoder:}} Given $w_1$, $w_2$, the transmitter generates $m_c$, $\tilde{m}_c=\{\tilde{m}_{c,1},\tilde{m}_{c,2}\}$ as in (\ref{eq:M_c}), (\ref{eq:M_c_tilde}), and sends  $\bx_c^n$, from $\setC_{c,n}$, which corresponds to $m_c$, $\tilde{m}_{c,1}$, $\tilde{m}_{c,2}$, i.e., $\bx_c^n\Big(m_c,\tilde{m}_{c,1}^{(1)},\tilde{m}_{c,2}^{(1)},\cdots,\tilde{m}_{c,1}^{(n\frac{\alpha}{2})},\tilde{m}_{c,2}^{(n\frac{\alpha}{2})}\Big)$. 

For the delivery phase, define $M_{\bd}$ and $\tilde{M}_{\bd}$ as in (\ref{eq:M_d}) and (\ref{eq:M_d_tilde}). $M_{\bd}$ represents the message to be securely transmitted during delivery no matter what the adversary's choice of $\alpha_2$ is. $\tilde{M}_{\bd}$ represents the randomization message.  Similar to cache placement, the transmitter further divides $\tilde{M}_{\bd,1}$, $\tilde{M}_{\bd,2}$ into sequences of independent bits, $\{\tilde{M}_{\bd,1}^{(1)},\cdots,\tilde{M}_{\bd,1}^{(n\frac{\alpha}{2})}\}$, $\{\tilde{M}_{\bd,2}^{(1)},\cdots,\tilde{M}_{\bd,2}^{(n\frac{\alpha}{2})}\}$, and generates $\bX_{\bd}^n$ as follows.

\begin{figure*}
\centering
\includegraphics[scale=0.63]{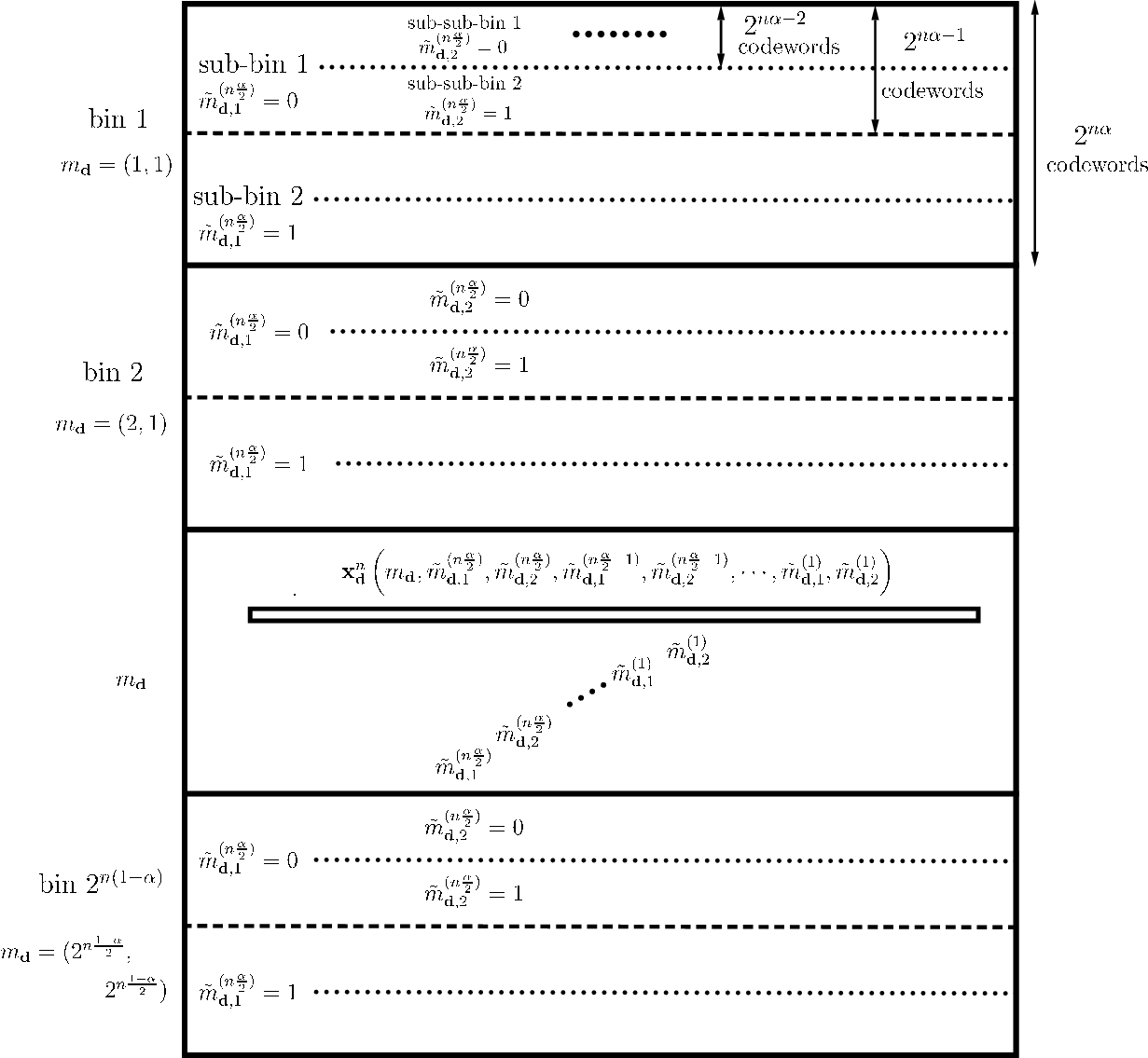}
\vspace{0.8cm} \caption{Codebook construction for the delivery phase, $\setC_{\bd,n}$.}
\label{fig:code_construction_2}
\end{figure*}

{\it{Delivery Codebook Generation:}} Let $m_{\bd}$, $\tilde{m}_{\bd,1}=\big\{\tilde{m}_{\bd,1}^{(1)},\cdots,\tilde{m}_{\bd,1}^{(n\frac{\alpha}{2})}\big\}$, $\tilde{m}_{\bd,2}=\big\{\tilde{m}_{\bd,2}^{(1)},\cdots,\tilde{m}_{\bd,2}^{(n\frac{\alpha}{2})}\big\}$ be the realizations of $M_{\bd}$, $\tilde{M}_{\bd,1}$, $\tilde{M}_{\bd,2}$. We construct $\setC_{\bd,n}$, from which $\bX_{\bd}^n$ is drawn, in a similar fashion as $\setC_{c,n}$, but with a reversed indexing of the sub-bins. We randomly and independently divide all the $2^{n}$ binary sequences into $2^{n(1-\alpha)}$ bins, indexed by $m_{\bd} \in [1:2^{n\frac{1-\alpha}{2}}]^2$, and each contains $2^{n\alpha}$ codewords. We further randomly and independently divide each bin $m_{\bd}$ into two sub-bins, indexed by $\tilde{m}_{\bd,1}^{(n\frac{\alpha}{2})}$, and each contains $2^{n\alpha-1}$ codewords. The process continues, going in reverse order over $\tilde{m}_{\bd,2}^{(n\frac{\alpha}{2})},\tilde{m}_{\bd,1}^{(n\frac{\alpha}{2}-1)}$, $\tilde{m}_{\bd,2}^{(n\frac{\alpha}{2}-1)}$, $\cdots$, $\tilde{m}_{\bd,1}^{(1)}$, until the remaining two codewords, after each sequence of divisions, are indexed by $\tilde{m}_{\bd,2}^{(1)}$. The codebook $\setC_{\bd,n}$ is shown in Fig. \ref{fig:code_construction_2}. 

{\it{Delivery Encoder:}} Given $w_1$, $w_2$ and $\bd=(d_1,d_2)$, the transmitter generates $m_{\bd}$, $\tilde{m}_{\bd}=\{\tilde{m}_{\bd,1},\tilde{m}_{\bd,2}\}$ as in (\ref{eq:M_d}), (\ref{eq:M_d_tilde}). The transmitter sends $\bx_{\bd}^n$, from $\setC_{\bd,n}$, which corresponds to $m_{\bd}$, $\tilde{m}_{\bd,1}$, and $\tilde{m}_{\bd,2}$, i.e., $\bx_{\bd}^n\big(m_{\bd},\tilde{m}_{\bd,1}^{(n\frac{\alpha}{2})},\tilde{m}_{\bd,2}^{(n\frac{\alpha}{2})},\cdots,\tilde{m}_{\bd,1}^{(1)},\tilde{m}_{\bd,2}^{(1)}\big)$. 

{\it{Decoding:}} Using $\bX_c^n$, receiver $j$, $j=1,2,$  recovers $M_{c,j}$, $\tilde{M}_{c,j}$, and stores them in its cache. For $j=1,2,$ the combined size of $M_{c,j}$ and $\tilde{M}_{c,j}$ does not exceed $\frac{n}{2}$ bits. Using $\bX_{\bd}^n$, both receivers recover $M_{\bd}$, $\tilde{M}_{\bd}$. Using $M_{\bd}$, $\tilde{M}_{\bd}$, $M_{c,j}$, $\tilde{M}_{c,j}$, and for $n$ sufficiently large, receiver $j$ correctly decodes $W_{d_j}$. 

{\it{Security Analysis:}} Let us first slightly modify the construction above as follows. Recall that $\{\epsilon_n\}_{n\geq 1}$ is a sequence of positive real numbers such that $\epsilon_n \rightarrow 0$ as $n \rightarrow\infty$. Define 
\begin{align}
\label{eq:alpha_epsilon}
\alpha_{\epsilon}=\alpha+2\epsilon_n,\;
\alpha_{1,\epsilon}=\alpha_1+\epsilon_n,\;
\alpha_{2,\epsilon}=\alpha_2+\epsilon_n.
\end{align}
That is, $\alpha_{1,\epsilon}+\alpha_{2,\epsilon}=\alpha_{\epsilon}$. We increase the sizes of $K_1$, $K_2$, into $\frac{n\alpha_{\epsilon}}{2}$ bits, from $n\frac{\alpha}{2}$, and zero-pad the bit strings of $W_{d_1,s}$, $W_{d_2,s}$, accordingly. We also decrease the sizes of $W_l^{(1)}$, $W_l^{(2)}$, $l=1,2,$ to $n\frac{1-\alpha_{\epsilon}}{2}$ bits, instead of $n\frac{1-\alpha}{2}$. We assume that $\frac{n\alpha_{\epsilon}}{2}$, $\frac{n\alpha_{1,\epsilon}}{2}$ are integers; as minor modifications can be adopted otherwise. 
 
Fix $S_1,S_2\subseteq [1:n]$. For the corresponding (fixed) values of $\alpha_1$, $\alpha_2$, the cache placement codebook $\setC_{c,n}$ can be viewed as a wiretap code with $2^{n(1-\alpha_{1,\epsilon})}$ bins. Each bin is indexed by the message
\begin{align}
\label{eq:w_c}
w_c=\big(m_c,\tilde{m}_{c,1}^{(1)},\tilde{m}_{c,2}^{(1)},\tilde{m}_{c,1}^{(2)},\tilde{m}_{c,2}^{(2)},\cdots,\tilde{m}_{c,1}^{(n\frac{\alpha_{2,\epsilon}}{2})},\tilde{m}_{c,2}^{(n\frac{\alpha_{2,\epsilon}}{2})}\big).
\end{align}
Each bin $w_c$ contains $2^{n\alpha_{1,\epsilon}}$ binary codewords which are indexed by the randomization message
\begin{align}
\label{eq:w_c_tilde}
\nonumber &\tilde{w}_c=\big(\tilde{m}_{c,1}^{(n\frac{\alpha_{2,\epsilon}}{2}+1)},\tilde{m}_{c,2}^{(n\frac{\alpha_{2,\epsilon}}{2}+1)},\tilde{m}_{c,1}^{(n\frac{\alpha_{2,\epsilon}}{2}+2)},\tilde{m}_{c,2}^{(n\frac{\alpha_{2,\epsilon}}{2}+2)},\\
&\qquad \quad \cdots,\tilde{m}_{c,1}^{(n\frac{\alpha_{\epsilon}}{2})},\tilde{m}_{c,2}^{(n\frac{\alpha_{\epsilon}}{2})}\big).
\end{align}

Similarly, $\setC_{\bd,n}$ can be seen as a wiretap code with $2^{n(1-\alpha_{2,\epsilon})}$ bins, each is indexed by the message
\begin{align}
\label{eq:w_d}
\nonumber & w_{\bd}=\big(m_{\bd},\tilde{m}_{\bd,1}^{(n\frac{\alpha_{\epsilon}}{2})},\tilde{m}_{\bd,2}^{(n\frac{\alpha_{\epsilon}}{2})},\tilde{m}_{\bd,1}^{(n\frac{\alpha_{\epsilon}}{2}-1)},\tilde{m}_{\bd,2}^{(n\frac{\alpha_{\epsilon}}{2}-1)},\\
&\qquad\qquad \cdots,\tilde{m}_{\bd,1}^{(n\frac{\alpha_{2,\epsilon}}{2}+1)},\tilde{m}_{\bd,2}^{(n\frac{\alpha_{2,\epsilon}}{2}+1)}\big).
\end{align}
Each bin $w_{\bd}$ contains $2^{n\alpha_{2,\epsilon}}$ codewords, indexed by the randomization message
\begin{align}
\label{eq:w_d_tilde}
\nonumber &\tilde{w}_{\bd}=\big(\tilde{m}_{\bd,1}^{(n\frac{\alpha_{2,\epsilon}}{2})},\tilde{m}_{\bd,2}^{(n\frac{\alpha_{2,\epsilon}}{2})},\tilde{m}_{\bd,1}^{(n\frac{\alpha_{2,\epsilon}}{2}-1)},\tilde{m}_{\bd,2}^{(n\frac{\alpha_{2,\epsilon}}{2}-1)},\\
&\qquad\qquad\cdots,\tilde{m}_{\bd,1}^{(1)},\tilde{m}_{\bd,2}^{(1)}\big). 
\end{align}

Let $\left\{\setB_{w_c}:w_c\in[1:2^{n(1-\alpha_{1,\epsilon})}]\right\}$,$\{\setB_{w_{\bd}}:w_{\bd}\in[1:2^{n(1-\alpha_{2,\epsilon})}]\}$ denote the partition (bins) of $\setC_{c,n}$, $\setC_{\bd,n}$, which correspond to $w_c$, $w_{\bd}$ in (\ref{eq:w_c}), (\ref{eq:w_d}), respectively. Let $\bx^{2n}\triangleq(\bx_{c}^n,\bx_{\bd}^n)$ denote the concatenation of the two length-$n$ binary codewords $\bx_c^n$, $\bx_{\bd}^n$.  Define the Cartesian product of the bins $\setB_{w_c}$ and $\setB_{w_{\bd}}$, as
\begin{align}
\label{eq:partition_B_w}
&\setB_{w_c,w_{\bd}}\triangleq \left\{\bx^{2n}=(\bx_{c}^n,\bx_{\bd}^n):\bx_c^n\in\setB_{w_c},\;\bx_{\bd}^n\in \setB_{w_{\bd}}\right\}.
\end{align}
Since the partitioning of $\setC_{c,n}$ and $\setC_{\bd,n}$ is random, for every $w_c$, $w_{\bd}$; $\setB_{w_c,w_{\bd}}$ is a random codebook which results from the Cartesian product of the random bins $\setB_{w_c}$ and $\setB_{w_{\bd}}$. Recall that $\setB_{w_c}$ contains $2^{n\alpha_{1,\epsilon}}$ and $\setB_{w_{\bd}}$ contains $2^{n\alpha_{2,\epsilon}}$ length-$n$ binary codewords. Thus, the product $\setB_{w_c,w_{\bd}}$ contains $2^{n\alpha_{\epsilon}}$ length-$2n$ binary codewords. Let $\{W_{d_l,s}^{(1)},\cdots,W_{d_l,s}^{(n\frac{\alpha_{\epsilon}}{2})}\}$ and $\{K_l^{(1)},\cdots,K_l^{(n\frac{\alpha_{\epsilon}}{2})}\}$ denote the bit strings of $W_{d_l,s}$ and $K_l$, $l=1,2$. In addition, for notational simplicity, define 
\begin{align}
\label{eq:Ws-1}
\bW_{s}^{(1)}&=\{W_{d_1,s}^{(1)},W_{d_2,s}^{(1)},\cdots,W_{d_1,s}^{(n\frac{\alpha_{2,\epsilon}}{2})},W_{d_2,s}^{(n\frac{\alpha_{2,\epsilon}}{2})}\}\\
\label{eq:Ws-2}
\bW_{s}^{(2)}&=\{W_{d_1,s}^{(n\frac{\alpha_{2,\epsilon}}{2}+1)},W_{d_2,s}^{(n\frac{\alpha_{2,\epsilon}}{2}+1)},\cdots,W_{d_1,s}^{(n\frac{\alpha_{\epsilon}}{2})},W_{d_2,s}^{(n\frac{\alpha_{\epsilon}}{2})}\}\\
\label{eq:K-1}
\bK^{(1)}&=\{K_1^{(1)},K_2^{(1)},\cdots,K_1^{(n\frac{\alpha_{2,\epsilon}}{2})},K_2^{(n\frac{\alpha_{2,\epsilon}}{2})}\}\\
\label{eq:K-2}
\bK^{(2)}&=\{K_1^{(n\frac{\alpha_{2,\epsilon}}{2}+1)},K_2^{(n\frac{\alpha_{2,\epsilon}}{2}+1)},\cdots,K_1^{(n\frac{\alpha_{\epsilon}}{2})},K_2^{(n\frac{\alpha_{\epsilon}}{2})}\}\\
\label{eq:K-Ws-1}
\bW_{\oplus\bK}^{(1)}&=\{W_{d_1,s}^{(i)}\oplus K_1^{(i)},W_{d_2,s}^{(i)}\oplus K_2^{(i)}:\; i=1,2,\cdots,n\frac{\alpha_{2,\epsilon}}{2}\}\\
\label{eq:K-Ws-2}
\nonumber \bW_{\oplus\bK}^{(2)}&=\{W_{d_1,s}^{(i)}\oplus K_1^{(i)},W_{d_2,s}^{(i)}\oplus K_2^{(i)}:\;  \\
&\qquad \quad i=n\frac{\alpha_{2,\epsilon}}{2}+1, n\frac{\alpha_{2,\epsilon}}{2}+2,\cdots,n\frac{\alpha_{\epsilon}}{2}\}.
\end{align}
Let $W_c$, $\tilde{W}_c$, $W_{\bd}$, and $\tilde{W}_{\bd}$ denote the random variables that correspond to the realizations defined in (\ref{eq:w_c})--(\ref{eq:w_d_tilde}). Using (\ref{eq:M_c})--(\ref{eq:M_d_tilde}), (\ref{eq:w_c})--(\ref{eq:w_d_tilde}), and (\ref{eq:K-1})--(\ref{eq:K-Ws-2}), we have
\begin{align}
\label{eq:Wc-Wc_tilde}
\nonumber &W_c=\{M_c,\bK^{(1)}\}=\{W_1^{(1)}\oplus W_2^{(1)},W_1^{(2)}\oplus W_2^{(2)},\bK^{(1)}\},\\ & \qquad \qquad \tilde{W}_c=\bK^{(2)}\\
\label{eq:Wd-Wd_tilde}
&W_{\bd}=\{M_{\bd},\bW_{\oplus\bK}^{(2)}\}=\{W_{d_1}^{(2)},W_{d_2}^{(1)},\bW_{\oplus\bK}^{(2)}\},\;\; \tilde{W}_{\bd}=\bW_{\oplus\bK}^{(1)}.
\end{align}
Notice that $\tilde{W}_c$ and $\tilde{W}_{\bd}$ are independent, and each is uniformly distributed. $\{\tilde{W}_c,\tilde{W}_{\bd}\}$ is thus jointly uniform. In addition, $\{\tilde{W}_c,\tilde{W}_{\bd}\}$ is independent from $\{W_c,W_{\bd}\}$. Thus, we can apply the analysis in \cite[($94$)-($103$)]{goldfeld2015semantic} to show that, for every $S_1$, $S_2$, $w_c$, and $w_{\bd}$, and every $\epsilon>0$, there exists $\gamma(\epsilon)>0$ such that
\begin{align}
\label{eq:Proof_1_1}
\nonumber \mathbb{P}_{\setB_{w_c,w_{\bd}}}\big(&\mathbb{D}\big(P_{\bZ_{S_1}^n\bZ_{S_2}^n\big|W_c=w_c,W_{\bd}=w_{\bd}}||P_{\bZ_{S_1}^n\bZ_{S_2}^n}\big)>\epsilon\big)\\
&\qquad\quad \leq \exp(-e^{n\gamma(\epsilon)}).
\end{align}
$P_{\bZ_{S_1}^n\bZ_{S_2}^n\big|W_c=w_c,W_{\bd}=w_{\bd}}$ is the induced distribution at the adversary when $\bx_c^n(w_c,\tilde{w}_c)$, $\bx_{\bd}^n(w_{\bd},\tilde{w}_{\bd})$ are the transmitted codewords over placement and delivery. $P_{\bZ_{S_1}^n\bZ_{S_2}^n}$ is the output distribution at the adversary when the placement and delivery codewords $\bX_c^n$, $\bX_{\bd}^n$ are drawn independently and identically at random according to the input distribution, which is uniform over $\{0,1\}$. (\ref{eq:Proof_1_1}) states that the probability that these two distributions are not equal converges to zero doubly-exponentially fast with the block-length $n$.

The number of the messages $\{w_c,w_{\bd}\}$ is $2^{n(2-\alpha_{\epsilon})}$, and the number of possible choices for the subsets $S_1$, $S_2$, is $\binom{2n}{\alpha n}<2^{2n}$. The combined number of the messages and the subsets is at most exponential in $n$. Using (\ref{eq:Proof_1_1}) and the union bound, as in \cite{goldfeld2015semantic,nafea2018new}, we have 
\begin{align}
\label{eq:Proof_1_2}
\limitn \max_{S_1,S_2}I(W_c,W_{\bd};\bZ_{S_1}^n,\bZ_{S_2}^n)=0.
\end{align}

We also have, for any $\bd=(d_1,d_2)$, $d_1,d_2\in\{1,2\}$, 
\begin{align}
\label{eq:Proof_1_3_0}
\nonumber & I(W_1,W_2;\bZ_{S_1}^n,\bZ_{S_2}^n)\\
&=I(W_1^{(1)},W_1^{(2)},W_2^{(1)},W_2^{(2)},W_{1,s},W_{2,s};\bZ_{S_1}^n,\bZ_{S_2}^n)\\
\label{eq:Proof_1_3}
&= I(W_1^{(1)},W_1^{(2)},W_2^{(1)},W_2^{(2)},\bW_{s}^{(1)},\bW_{s}^{(2)};\bZ_{S_1}^n,\bZ_{S_2}^n)\\
\label{eq:Proof_1_4}
\nonumber &=I(W_1^{(1)}\oplus W_2^{(1)},W_1^{(2)}\oplus W_2^{(2)},W_{d_1}^{(2)},W_{d_2}^{(1)},\\
&\qquad\qquad\qquad \qquad\qquad \bW_{s}^{(1)},\bW_{s}^{(2)};\bZ_{S_1}^n,\bZ_{S_2}^n)\\
\label{eq:Proof_1_5}
&=I(M_c,M_{\bd},\bW_{s}^{(1)},\bW_{s}^{(2)};\bZ_{S_1}^n,\bZ_{S_2}^n)\\
\label{eq:Proof_1_6}
&\leq I(M_c,M_{\bd},\bW_{s}^{(1)},\bW_{\oplus\bK}^{(2)};\bZ_{S_1}^n,\bZ_{S_2}^n)\\
\label{eq:Proof_1_7}
&=I(M_c,\bW_s^{(1)},W_{\bd};\bZ_{S_1}^n,\bZ_{S_2}^n)\\
\label{eq:Proof_1_8}
&= H(\bZ_{S_1}^n,\bZ_{S_2}^n)-H(\bZ_{S_1}^n,\bZ_{S_2}^n\Big|M_c,\bW_s^{(1)},W_{\bd}),
\end{align}
where (\ref{eq:Proof_1_3}) follows because $\bZ_{S_1}^n$, $\bZ_{S_2}^n$ depend only on $\{W_1^{(1)},W_1^{(2)},W_2^{(1)},W_2^{(2)},W_{d_1,s},W_{d_2,s}\}$, and by using (\ref{eq:Ws-1}), (\ref{eq:Ws-2}); (\ref{eq:Proof_1_4}) follows because there is a bijection between $\{W_1^{(1)},W_1^{(2)},W_2^{(1)},W_2^{(2)}\}$ and $\{W_1^{(1)}\oplus W_2^{(1)},W_1^{(2)}\oplus W_2^{(2)},W_{d_1}^{(2)},W_{d_2}^{(1)}\}$; (\ref{eq:Proof_1_5}) follows from (\ref{eq:M_c}), (\ref{eq:M_d}); (\ref{eq:Proof_1_6}) follows due to the Markov chain $\bW_s^{(2)}-\{M_c,M_{\bd},\bW_s^{(1)},\bW_{\oplus\bK}^{(2)}\}-\{\bZ_{S_1}^n,\bZ_{S_2}^n\}$, and the data processing inequality. This Markov chain holds because $\{M_c,M_{\bd},\bW_s^{(1)}\}$ are independent from $\{\bW_s^{(2)},\bK^{(2)}\}$, and only the encrypted information $\bW_{\oplus\bK}^{(2)}$ is transmitted. (\ref{eq:Proof_1_7}) follows from (\ref{eq:Wd-Wd_tilde}). The second term on the RHS of (\ref{eq:Proof_1_8}) is lower bounded as
\begin{align}
\label{eq:Proof_1_9}
\nonumber & H(\bZ_{S_1}^n,\bZ_{S_2}^n\big|M_c,\bW_s^{(1)},W_{\bd})\\
&=H(\bZ_{S_1}^n,\bZ_{S_2}^n,\bW_s^{(1)}\big|M_c,W_{\bd})-H(\bW_s^{(1)}\big|M_c,W_{\bd})\\
\label{eq:Proof_1_10}
\nonumber&=H(\bZ_{S_1}^n,\bZ_{S_2}^n,\bW_s^{(1)},\bW_{\oplus K}^{(1)}\big|M_c,W_{\bd})\\
&\;\; -H(\bW_{\oplus K}^{(1)}\big|M_c,W_{\bd},\bW_s^{(1)},\bZ_{S_1}^n,\bZ_{S_2}^n)-H(\bW_s^{(1)})\\
\label{eq:Proof_1_11}
\nonumber&=H(\bZ_{S_1}^n,\bZ_{S_2}^n,\bK^{(1)},\bW_{\oplus K}^{(1)}\big|M_c,W_{\bd})\\
&\;\; -H(\bK^{(1)}\big|M_c,W_{\bd},\bW_s^{(1)},\bZ_{S_1}^n,\bZ_{S_2}^n)-H(\bW_s^{(1)})\\
\label{eq:Proof_1_12}
&\geq H(\bZ_{S_1}^n,\bZ_{S_2}^n,\bK^{(1)},\bW_{\oplus K}^{(1)}\big|M_c,W_{\bd})-H(\bW_s^{(1)})-\epsilon'_n\\
\label{eq:Proof_1_13}
\nonumber &\geq H(\bK^{(1)}\big|M_c,W_{\bd})+H(\bZ_{S_1}^n,\bZ_{S_2}^n\big|M_c,\bK^{(1)},W_{\bd})\\
&\qquad\qquad\qquad\qquad \qquad -H(\bW_s^{(1)})-\epsilon'_n\\
\label{eq:Proof_1_14}
&=H(\bZ_{S_1}^n,\bZ_{S_2}^n\big|W_c,W_{\bd})+ H(\bK^{(1)})-H(\bW_s^{(1)})-\epsilon'_n\\
\label{eq:Proof_1_15}
&= H(\bZ_{S_1}^n,\bZ_{S_2}^n\big|W_c,W_{\bd})-\epsilon'_n,
\end{align}
where $\epsilon'_n \rightarrow 0$ as $n \rightarrow \infty$; (\ref{eq:Proof_1_10}) follows since $\bW_s^{(1)}$ is independent from $\{M_c,W_{\bd}\}$; (\ref{eq:Proof_1_11}) follows because there is a bijection between $\{\bW_s^{(1)},\bW_{\oplus K}^{(1)}\}$ and $\{\bK^{(1)},\bW_{\oplus K}^{(1)}\}$; (\ref{eq:Proof_1_14}) follows from (\ref{eq:Wc-Wc_tilde}), and since $\bK^{(1)}$ is independent from $\{M_c,W_{\bd}\}$; (\ref{eq:Proof_1_15}) follows since $\bK^{(1)}$, $\bW_s^{(1)}$ are independent and identically distributed. 

The inequality in (\ref{eq:Proof_1_12}) follows since, given $\{M_c$,$\bW_s^{(1)}$,$W_{\bd}\}$, and for $n$ sufficiently large, the adversary can decode $\bK^{(1)}$ using its observations $\bZ_{S_1}^n$, $\bZ_{S_2}^n$. In particular, $\{M_c,\bW_s^{(1)},W_{\bd}\}$ determine a partition of the codebook into bins, each of which contains $2^{n\alpha_{\epsilon}}$ codewords. For $n$ large enough, and given the values of $M_c$, $\bW_s^{(1)}$, $W_{\bd}$, i.e., the bin index, the adversary can determine the codeword index inside the bin, and hence decode $\bK^{(1)}$. We conclude that, $H(\bK^{(1)}\big|M_c,W_{\bd},\bW_s^{(1)},\bZ_{S_1}^n,\bZ_{S_2}^n)\leq \epsilon'_n$, where $\epsilon'_n \rightarrow 0$ as $n \rightarrow\infty$. 

Substituting (\ref{eq:Proof_1_15}) in (\ref{eq:Proof_1_8}) gives
\begin{align}
\label{eq:Proof_1_16}
I(W_1,W_2;\bZ_{S_1}^n,\bZ_{S_2}^n)\leq I(W_c,W_{\bd};\bZ_{S_1}^n,\bZ_{S_2}^n)+\epsilon'_n.
\end{align}
Using (\ref{eq:Proof_1_2}) and (\ref{eq:Proof_1_16}), the secrecy constraint in (\ref{eq:secrecy}) is satisfied. Since $\epsilon_n \rightarrow 0$ as $n \rightarrow\infty$, we conclude that, the achievable strong secrecy file rate is 
\begin{align}
\label{eq:Proof_1_18}
R_s(\alpha)=2\times \frac{1-\alpha}{2}+ \frac{\alpha}{2}=1-\frac{\alpha}{2}.
\end{align}

\begin{remark}
Although the codebooks $\setC_{c,n}$, $\setC_{\bd,n}$ are designed and generated disjointly, in the security analysis, we considered the Cartesian products of the individual bins of the two codebooks. We were able to do so since the input distributions for generating the codebooks are identical, i.e., uniform binary.
\end{remark}
\subsection{Achievability for $\alpha\in[1,2]$:}\label{Achievability_2}
For $\alpha\in [1,2]$, we adapt the scheme in Section \ref{Achievability_1} as follows: $W_1$, $W_2$ are uniform over $[1:2^{n\frac{2-\alpha_{\epsilon}}{2}}]$; $\alpha_{\epsilon}$ is as in (\ref{eq:alpha_epsilon}). The transmitter (i) generates the independent keys $K_1$, $K_2$, uniform over $[1:2^{n\frac{2-\alpha_{\epsilon}}{2}}]$ and independent from $W_1$, $W_2$; (ii) generates the independent randomization messages $\tilde{W},\tilde{W}_K$, uniform over $[1:2^{n(\alpha_{\epsilon}-1)}]$ and independent from $W_1$, $W_2$, $K_1$, $K_2$. The messages for placement at receivers $1, 2$ are
\begin{align}
\label{eq:Mc1-Mc2}
M_{c,1}=K_1,\qquad M_{c,2}=K_2,
\end{align}
i.e., receiver $j$ stores the key $K_j$ in its cache. The message to be securely transmitted during delivery is
\begin{align}
\label{eq:Md1-Md2}
M_{\bd}=\{M_{\bd,1},M_{\bd,2}\};M_{\bd,1}=W_{d_1}\oplus K_1, M_{\bd,2}=W_{d_2}\oplus K_2.
\end{align}

Note that, for $\alpha\in[1,2]$, the adversary can see all symbols in at least one of the phases. Therefore, unlike Section \ref{Achievability_1}, we cannot utilize randomization messages, $\tilde{W}$, $\tilde{W}_K$, to carry any information; only keys are stored in the cache memories of the receivers. In addition, the placement and delivery codebooks for this case have a different embedding structure than for $\alpha\in(0,1)$ in Section \ref{Achievability_1}. In particular, the embedding here is performed on the bits of the messages $M_c$, $M_{\bd}$, while the embedding in Section \ref{Achievability_1} is performed on the bits of the randomization messages $\tilde{M}_c$, $\tilde{M}_{\bd}$. Let $\{W_{d_l}^{(1)},\cdots,W_{d_l}^{(n\frac{2-\alpha_{\epsilon}}{2})}\}$, $\{K_l^{(1)},\cdots,K_l^{(n\frac{2-\alpha_{\epsilon}}{2})}\}$, $\{M_{\bd,l}^{(1)},\cdots,M_{\bd,l}^{(n\frac{2-\alpha_{\epsilon}}{2})}\}$ denote the bit strings of $W_{d_l}$, $K_l$, $M_{\bd,l}$, $l=1,2$. 

{\it{Cache Placement Codebook:}} The codebook $\setC_{c,n}$ is generated as follows: The transmitter randomly and independently divides all the $2^n$ length-$n$ binary sequences into $2$ bins, indexed by $K_1^{(1)}$, and each contains $2^{n-1}$ codewords. These bins are further randomly and independently divided into two sub-bins, indexed by $K_2^{(1)}$, and each contains $2^{n-2}$ codewords. The process continues, going over $K_1^{(2)}$, $K_2^{(2)}$, $\cdots$, $K_1^{(n\frac{2-\alpha_{\epsilon}}{2})}$, $K_2^{(n\frac{2-\alpha_{\epsilon}}{2})}$, until the remaining $2^{n(\alpha_{\epsilon}-1)}$ codewords, after each sequence of divisions, are indexed by $\tilde{W}_{K}$. 

{\it{Cache Encoder:}} The transmitter sends $\bold{X}_c^n$ which corresponds to the keys $K_1$, $K_2$, and the randomization message $\tilde{W}_K$, i.e., $\bold{X}_c^n\Big(K_1^{(1)},K_2^{(1)},\cdots,K_1^{(n\frac{2-\alpha_{\epsilon}}{2})}, K_2^{(n\frac{2-\alpha_{\epsilon}}{2})},\tilde{W}_K\Big)$.

{\it{Delivery Codebook:}} The codebook $\setC_{\bd,n}$ is generated as follows. The transmitter randomly and independently divides all the $2^n$ length-$n$ binary sequences into two bins, indexed by $M_{\bd,1}^{(n\frac{2-\alpha_{\epsilon}}{2})}$, and each contains $2^{n-1}$ codewords. These bins are further randomly and independently divided into two sub-bins, indexed by $M_{\bd,2}^{(n\frac{2-\alpha_{\epsilon}}{2})}$, and each contains $2^{n-2}$ codewords. The process continues, going in reverse order over $M_{\bd,1}^{(n\frac{2-\alpha_{\epsilon}}{2}-1)}$, $M_{\bd,2}^{(n\frac{2-\alpha_{\epsilon}}{2}-1)}$, $\cdots$, $M_{\bd,1}^{(1)}$, $M_{\bd,2}^{(1)}$, until the remaining $2^{n(\alpha_{\epsilon}-1)}$ codewords, after each sequence of divisions, are indexed by the randomization message $\tilde{W}$. 

{\it{Delivery Encoder:}} Given $W_1$, $W_2$, $K_1$, $K_2$, $\tilde{W}$, $\bd=(d_1,d_2)$, the transmitter forms $M_{\bd,1}$, $M_{\bd,2}$, as in (\ref{eq:Md1-Md2}) and sends $\bold{X}_{\bd}^n$ which corresponds to $M_{\bd,1}$, $M_{\bd,2}$, $\tilde{W}$, i.e., $\bold{X}_{\bd}^n\Big(M_{\bd,1}^{(n\frac{2-\alpha_{\epsilon}}{2})},M_{\bd,2}^{(n\frac{2-\alpha_{\epsilon}}{2})},\cdots,M_{\bd,1}^{(1)}, M_{\bd,2}^{(1)},\tilde{W}\Big)$.

{\it{Decoding:}} Using $\bX_c^n$, receiver $j=1,2,$ recovers $M_{c,j}=K_j$ and stores it in its cache. Using $\bX_{\bd}^n$, both receivers recover $M_{\bd}=\{M_{\bd,1},M_{\bd,2}\}$. Using $M_{\bd,j}$, $K_j$, and for $n$ large enough, receiver $j$ recovers $W_{d_j}$. 

{\it{Security Analysis:}} Fix $S_1$, $S_2$. Recall that $\alpha_1,\alpha_2\leq 1$. Since $\alpha\geq 1$, $\alpha_1,\alpha_2\geq \alpha -1$. If $\alpha_1=1$, then $\alpha_2=\alpha-1$, and vice versa. In addition, notice that $1-\alpha_1,1-\alpha_2\leq 2-\alpha$. As in Section \ref{Achievability_1}, for a fixed value of $\alpha_1$, the codebook $\setC_{c,n}$ is a wiretap code with $2^{n(1-\alpha_{1,\epsilon})}$ bins, indexed by 
\begin{align}
\label{eq:Wc}
W_c=\big(K_1^{(1)},K_2^{(1)},\cdots,K_1^{(n\frac{1-\alpha_{1,\epsilon}}{2})},K_2^{(n\frac{1-\alpha_{1,\epsilon}}{2})}\big).
\end{align}
Each bin $W_c$ contains $2^{n\alpha_{1,\epsilon}}$ binary codewords, indexed by
\begin{align}
\label{eq:Wctilde}
\nonumber &\tilde{W}_c=\big(K_1^{(n\frac{1-\alpha_{1,\epsilon}}{2}+1)},K_2^{(n\frac{1-\alpha_{1,\epsilon}}{2}+1)},\cdots,\\
&\qquad\qquad \qquad \qquad K_1^{(n\frac{2-\alpha_{\epsilon}}{2})},K_2^{(n\frac{2-\alpha_{\epsilon}}{2})},\tilde{W}_K\big).
\end{align}
Similarly, for a fixed value of $\alpha_2$, $\setC_{\bd,n}$ is a wiretap code with $2^{n(1-\alpha_{2,\epsilon})}$ bins, each is indexed by 
\begin{align}
\label{eq:Wd}
\nonumber & W_{\bd}=\big(\tilde{M}_{\bd,1}^{(n\frac{2-\alpha_{\epsilon}}{2})},\tilde{M}_{\bd,2}^{(n\frac{2-\alpha_{\epsilon}}{2})},\cdots,\\
&\qquad\qquad\qquad\qquad\tilde{M}_{\bd,1}^{(n\frac{1-\alpha_{1,\epsilon}}{2}+1)},\tilde{M}_{\bd,2}^{(n\frac{1-\alpha_{1,\epsilon}}{2}+1)}\big).
\end{align}
Each bin $W_{\bd}$ contains $2^{n\alpha_{2,\epsilon}}$ codewords, indexed by 
\begin{align}
\label{eq:Wdtilde}
\tilde{W}_{\bd}=\big(\tilde{M}_{\bd,1}^{(n\frac{1-\alpha_{1,\epsilon}}{2})},\tilde{M}_{\bd,2}^{(n\frac{1-\alpha_{1,\epsilon}}{2})},\cdots,\tilde{M}_{\bd,1}^{(1)},\tilde{M}_{\bd,2}^{(1)},\tilde{W}\big). 
\end{align}

Let us re-define $\bK^{(1)},\bK^{(2)},\bW_{\oplus\bK}^{(1)},$ and $\bW_{\oplus\bK}^{(2)}$ and define $\bW^{(1)}$ and $\bW^{(2)}$ as
\begin{align}
\label{eq:K1}
\bK^{(1)}&=\big\{K_1^{(i)},K_2^{(i)}:\; i=1,\cdots,n\frac{1-\alpha_{1,\epsilon}}{2}\big\}\\
\label{eq:K2}
\bK^{(2)}&=\big\{K_1^{(i)},K_2^{(i)}:\; i=n\frac{1-\alpha_{1,\epsilon}}{2}+1,\cdots,n\frac{2-\alpha_{\epsilon}}{2}\big\}\\
\label{eq:K-W-1}
\bW_{\oplus\bK}^{(1)}&=\big\{W_{d_1}^{(i)}\oplus K_1^{(i)},W_{d_2}^{(i)}\oplus K_2^{(i)}: i=1,\cdots,n\frac{1-\alpha_{1,\epsilon}}{2}\big\}\\
\label{eq:K-W-2}
\nonumber \bW_{\oplus\bK}^{(2)}&=\big\{W_{d_1}^{(i)}\oplus K_1^{(i)},W_{d_2}^{(i)}\oplus K_2^{(i)}:\\
&\qquad\qquad i=n\frac{1-\alpha_{1,\epsilon}}{2}+1,\cdots,n\frac{2-\alpha_{\epsilon}}{2}\big\}\\
\label{eq:W-1}
\bW^{(1)}&=\big\{W_{d_1}^{(i)},W_{d_2}^{(i)}:\; i=1,\cdots,n\frac{1-\alpha_{1,\epsilon}}{2}\big\}\\
\label{eq:W-2}
\bW^{(2)}&=\big\{W_{d_1}^{(i)},W_{d_2}^{(i)}:\; i=n\frac{1-\alpha_{1,\epsilon}}{2}+1,\cdots,n\frac{2-\alpha_{\epsilon}}{2}\big\}.
\end{align}
From (\ref{eq:Wc})-(\ref{eq:K-W-2}), we have
\begin{align}
\label{eq:Wc-Wct-Wd-Wdt}
\nonumber &W_c=\bK^{(1)},\quad \tilde{W}_c=\{\bK^{(2)},\tilde{W}_K\}, \\ & W_{\bd}=\bW_{\oplus\bK}^{(2)},\quad \tilde{W}_{\bd}=\{\bW_{\oplus\bK}^{(1)},\tilde{W}\}.
\end{align}
Similar to Section \ref{Achievability_1}, $\tilde{W}_c$, $\tilde{W}_{\bd}$, are independent and uniform, and hence $\{\tilde{W}_c,\tilde{W}_{\bd}\}$ is jointly uniform. Also, $\{\tilde{W}_c,\tilde{W}_{\bd}\}$ is independent from $\{W_c,W_{\bd}\}$. Thus, (\ref{eq:Proof_1_2}) is satisfied. For any $\bd=(d_1,d_2)$, we have
\begin{align}
\label{eq:Proof_2_2}
I&(W_1,W_2;\bZ_{S_1}^n,\bZ_{S_2}^n)=I(\bW^{(1)},\bW^{(2)};\bZ_{S_1}^n,\bZ_{S_2}^n)\\
\label{eq:Proof_2_3}
&\leq I(\bW^{(1)},\bW_{\oplus\bK}^{(2)};\bZ_{S_1}^n,\bZ_{S_2}^n)\\
\label{eq:Proof_2_4}
&=I(\bW^{(1)},W_{\bd};\bZ_{S_1}^n,\bZ_{S_2}^n)\\
\label{eq:Proof_2_5}
&= H(\bZ_{S_1}^n,\bZ_{S_2}^n)-H(\bZ_{S_1}^n,\bZ_{S_2}^n\big|\bW^{(1)},W_{\bd})\\
\label{eq:Proof_2_6}
&\leq H(\bZ_{S_1}^n,\bZ_{S_2}^n)-H(\bZ_{S_1}^n,\bZ_{S_2}^n\big|\bK^{(1)},W_{\bd})+\epsilon'_n\\
\label{eq:Proof_2_7}
&= I(W_c,W_{\bd};\bZ_{S_1}^n,\bZ_{S_2}^n)+\epsilon'_n,
\end{align}
where (\ref{eq:Proof_2_3}) follows due to the Markov chain $\bW^{(2)}-\{\bW^{(1)},\bW_{\oplus\bK}^{(2)}\}-\{\bZ_{S_1}^n,\bZ_{S_2}^n\}$; (\ref{eq:Proof_2_4}), (\ref{eq:Proof_2_7}) follow from (\ref{eq:Wc-Wct-Wd-Wdt}); (\ref{eq:Proof_2_6}) follows by using similar steps as in (\ref{eq:Proof_1_9})-(\ref{eq:Proof_1_15}). Using (\ref{eq:Proof_1_2}) and (\ref{eq:Proof_2_7}), the secrecy constraint in (\ref{eq:secrecy}) is satisfied. Since $\epsilon_n \rightarrow 0$ as $n \rightarrow\infty$, the achievable strong secrecy file rate is 
\begin{align}
\label{eq:Proof_2_9}
R_s(\alpha)= \frac{2-\alpha}{2}=1-\frac{\alpha}{2}.
\end{align}
This completes the proof for Theorem \ref{thm:Thm1}. 

\section{Proof of Theorem \ref{thm:Thm2}}\label{Proof_Thm2}
In this section, we extend the achievability scheme in Section \ref{Proof_Thm1} and provide a lower bound on the strong secrecy file rate when $D>2$. The demand vector is $\bd=(d_1,d_2)$, where $d_1,d_2\in[1:D]$. As in Section \ref{Proof_Thm1}, we divide the proof into two cases for the ranges $\alpha\in(0,1)$ and $\alpha\in[1,2]$. 

\subsection{$\alpha\in[1,2]$}\label{Proof_Thm2_1}
For $\alpha\in[1,2]$, we use the same scheme as in Section \ref{Achievability_2}. For this range of $\alpha$, only the keys $K_1$, $K_2$, are transmitted during placement, and stored in receivers $1$ and $2$ cache memories. That is, no information messages are stored in the caches, and the user demands are known during delivery. Hence, $D>2$ is immaterial for this range of $\alpha$. The achievable strong secrecy file rate is $1-\frac{\alpha}{2}$.

\subsection{$\alpha\in(0,1)$}\label{Proof_Thm2_2}
The achievability scheme for this case has the same channel coding structure as in Section \ref{Achievability_1}. The difference however lies in generating the messages to be securely communicated over cache placement and delivery phases, i.e., $M_c$, $M_{\bd}$. In particular, we use here uncoded placement for designing the cache contents, and a partially coded delivery transmission that is simultaneously useful for both receivers.  

The transmitter (i) divides $W_l$, $l\in[1:D]$, into the independent messages $\{W_l^{(1)},W_l^{(2)},W_{l,t},W_{l,s}\}$, where $W_l^{(1)}$, $W_l^{(2)}$ are uniform over $[1:2^{n\frac{1-\alpha_{\epsilon}}{2D}}]$, $\alpha_{\epsilon}$ is as in (\ref{eq:alpha_epsilon}), $W_{l,t}$ is uniform over $[1:2^{n\frac{(2D-1)(1-\alpha_{\epsilon})}{4D}}]$, and $W_{l,s}$ is uniform over $[1:2^{n\frac{\alpha_{\epsilon}}{2}}]$; (ii) generates the independent keys $K_1$, $K_2$, uniform over $[1:2^{n\frac{\alpha_{\epsilon}}{2}}]$ and independent from $W_{[1:D]}$. 

Let $M_c= \{M_{c,1},M_{c,2}\}$. Unlike (\ref{eq:M_c}), we use here {\it{uncoded placement}} for designing $M_{c,1}$, $M_{c,2}$:
\begin{align}
\label{eq:Mc1_LB_1}
&M_{c,1}=\{W_1^{(1)},W_2^{(1)},\cdots,W_D^{(1)}\},\\
\label{eq:Mc2_LB_1}
&M_{c,2}=\{W_1^{(2)},W_2^{(2)},\cdots,W_D^{(2)}\}.
\end{align}
The randomization message $\tilde{M}_c$ is identical to (\ref{eq:M_c_tilde}). Receiver $j$ stores $M_{c,j}$, $\tilde{M}_{c,j}$ in its cache memory. 

Unlike (\ref{eq:M_d}), we use {\it{partially coded}} delivery.  The message to be securely transmitted in delivery is  
\begin{align}
\label{eq:Md_LB_1}
M_{\bd}=\{W_{d_2}^{(1)}\oplus W_{d_1}^{(2)}, W_{d_1,t}, W_{d_2,t}\}.
\end{align}
Notice that we use the term {\it{partially coded}} since part of $M_{\bd}$ is coded, i.e., $W_{d_2}^{(1)}\oplus W_{d_1}^{(2)}$, while the other part is uncoded, i.e., $W_{d_1,t}, W_{d_2,t}$. The randomization message for delivery, $\tilde{M}_{\bd}$, is identical to (\ref{eq:M_d_tilde}).

\begin{remark}
Note that the sizes of $M_c$, $M_{\bd}$, $\tilde{M}_c$, and $\tilde{M}_{\bd}$ are the same as in Section \ref{Achievability_1}. In particular, the sizes of $\tilde{M}_c$ and $\tilde{M}_{\bd}$ are both $n\alpha_{\epsilon}$ bits. The size of $M_c$ is 
$2\times D\times n\frac{1-\alpha_{\epsilon}}{2D}=n(1-\alpha_{\epsilon})$ bits
and the size of $M_{\bd}$ is 
$
n\frac{1-\alpha_{\epsilon}}{2D}+2\times n\frac{(2D-1)(1-\alpha_{\epsilon})}{4D}=n(1-\alpha_{\epsilon})$ bits.
\end{remark}

{\it{Codebooks Generation and Encoders:}} For the messages $M_c$, $M_{\bd}$, $\tilde{M}_c$, $\tilde{M}_{\bd}$ defined above, the cache placement and delivery codebooks, $\setC_{c,n}$ and $\setC_{\bd,n}$, and the cache and delivery encoders, are designed in the same exact manner as in Section \ref{Achievability_1}, see Figures \ref{fig:code_construction_1} and \ref{fig:code_construction_2}.

{\it{Decoding:}} As in Section \ref{Achievability_1}, using $M_{\bd}$, $\tilde{M}_{\bd}$, $M_{c,j}$, $\tilde{M}_{c,j}$, and for $n$ sufficiently large, receiver $j$ correctly decodes $W_{d_{j}}$, $j=1,2$. 

{\it{Security analysis:}} 
Let $W_c$, $\tilde{W}_c$, $W_{\bd}$, $\tilde{W}_{\bd}$, be defined as in (\ref{eq:w_c})-(\ref{eq:w_d_tilde}), (\ref{eq:Wc-Wc_tilde}), (\ref{eq:Wd-Wd_tilde}). Once again, $\tilde{W}_c$ and $\tilde{W}_{\bd}$ are independent and uniform, and hence $\{\tilde{W}_c,\tilde{W}_{\bd}\}$ is jointly uniform. In addition, $\{W_c,W_{\bd}\}$ are independent from $\{\tilde{W}_c,\tilde{W}_{\bd}\}$. Thus, (\ref{eq:Proof_1_2}) holds for this case. For any $\bd=(d_1,d_2)$, 
\begin{align}
\label{eq:Proof_LB_4}
\nonumber &I(W_{[1:D]};\bZ_{S_1}^n,\bZ_{S_2}^n)\\
&=I(\{W_l^{(1)},W_l^{(2)},W_{l,t},W_{l,s}\}_{l=1}^D;\bZ_{S_1}^n,\bZ_{S_2}^n)\\
\label{eq:Proof_LB_5}
&\leq I(M_c,\{W_l^{(1)},W_l^{(2)},W_{l,t},W_{l,s}\}_{l=1}^D;\bZ_{S_1}^n,\bZ_{S_2}^n)\\
\label{eq:Proof_LB_6}
&\leq I(M_c,W_{d_2}^{(1)}\oplus W_{d_1}^{(2)},W_{d_1,t},W_{d_2,t},W_{d_1,s},W_{d_2,s};\bZ_{S_1}^n,\bZ_{S_2}^n)\\
\label{eq:Proof_LB_7}
&=I(M_c,M_{\bd},W_{d_1,s},W_{d_2,s};\bZ_{S_1}^n,\bZ_{S_2}^n)\\
\label{eq:Proof_LB_8}
&\leq I(W_c,W_{\bd};\bZ_{S_1}^n,\bZ_{S_2}^n)+\epsilon'_n,
\end{align}
(\ref{eq:Proof_LB_6}) follows form the Markov chain $W_{[1:D]}-\{M_c,W_{d_2}^{(1)}\oplus W_{d_1}^{(2)},W_{d_1,t},W_{d_2,t},W_{d_1,s},W_{d_2,s}\}-\{\bZ_{S_1}^n,\bZ_{S_2}^n\}$; (\ref{eq:Proof_LB_7}) follows from (\ref{eq:Md_LB_1}); (\ref{eq:Proof_LB_8}) follows using similar steps as in (\ref{eq:Proof_1_4})-(\ref{eq:Proof_1_15}). Using (\ref{eq:Proof_1_2}), (\ref{eq:Proof_LB_8}), the secrecy constraint in (\ref{eq:secrecy}) is satisfied. With $\epsilon_n \rightarrow 0$ as $n \rightarrow\infty$, the achievable strong secrecy file rate is 
\begin{align}
\label{eq:Proof_LB_10}
R_s(\alpha)&= \frac{(1-\alpha)}{D}+\frac{(2D-1)(1-\alpha)}{4D}+\frac{\alpha}{2}=\frac{1}{2}+\frac{3(1-\alpha)}{4D}.
\end{align}
This completes the proof for Theorem \ref{thm:Thm2}. 

\begin{remark}
For $D=2$, the achievable secrecy rate in (\ref{eq:Proof_LB_10}) is strictly smaller than the secrecy rate obtained by {\it{coded}} placement and {\it{uncoded}} delivery in Section \ref{Achievability_1}, i.e., $1-\frac{\alpha}{2}$.
\end{remark}

\begin{remark}
An achievability scheme which utilizes coded placement and uncoded delivery, as in Section \ref{Achievability_1}, achieves the same secure file rate as (\ref{eq:Proof_LB_10}) for $D=3$. However, this scheme achieves a strictly smaller secure file rate for $D\geq 4$. In this scheme, $W_l^{(1)}$, $W_l^{(2)}$ are uniform over $[1:2^{n\frac{1-\alpha_{\epsilon}}{2(D-1)}}]$. $W_{l,t}$ is uniform over $[1:2^{n\frac{(D-2)(1-\alpha_{\epsilon})}{2(D-1)}}]$. $W_{l,s}$, $K_1$, $K_2$, are uniform over $[1:2^{n\frac{\alpha_{\epsilon}}{2}}]$. $M_c=\{M_{c,1},M_{c,2}\}$, where $M_{c,1}=\{W_1^{(1)}\oplus W_2^{(1)}, W_2^{(1)}\oplus W_3^{(1)},\cdots,W_{D-1}^{(1)}\oplus W_D^{(1)}\}$
and $M_{c,2}=\{W_1^{(2)}\oplus W_2^{(2)}, W_2^{(2)}\oplus W_3^{(2)},\cdots,W_{D-1}^{(2)}\oplus W_D^{(2)}\}$. $M_{\bd}=\{W_{d_2}^{(1)}, W_{d_1}^{(2)}, W_{d_1,t}, W_{d_2,t}\}$. Without loss of generality, let $d_1<d_2$. For any $\bd=(d_1,d_2)$, using $M_{c,j}$, receiver $j$, can restore $W_{d_1}^{(j)}\oplus W_{d_2}^{(j)}$ by xor-ing $\big\{W_{d_1}^{(j)}\oplus W_{d_1+1}^{(j)}\big\}$, $\big\{W_{d_1+1}^{(j)}\oplus W_{d_1+2}^{(j)}\big\}$,$\cdots$, $\big\{W_{d_2-1}^{(j)}\oplus W_{d_2}^{(j)}\big\}$. The achievable strong secrecy file rate using this scheme is $R_s(\alpha)=\frac{1}{2}+\frac{1-\alpha}{2(D-1)}$.
\end{remark}

\section{Proof of Theorem \ref{thm:Thm3}}\label{Proof_Thm3}
When $\alpha\in[1,2]$, the upper bound on $R_s$ in Theorem \ref{thm:Thm3} for $D>2$ follows as in Section \ref{Converse}. It remains to prove the upper bound for $\alpha\in(0,1)$. The proof is divided into the three following steps.  

{\it{Step $1$:}} We upper bound $R_s$ by the secrecy capacity when the adversary is restricted to tap into the delivery phase only, denoted as $C_s^{\rm{Res}}$. That is, $C_s^{\rm{Res}}$ is the maximum achievable file rate when $\alpha_1=0$, $\alpha_2=\alpha$. Restricting the adversary to only tap into the delivery phase cannot decrease the secrecy capacity, i.e., $R_s\leq C_s^{\rm{Res}}$, since this setting is included in its feasible strategy space. Cache placement is secure, and each receiver has a secure cache memory of size $\frac{n}{2}$ bits.

{\it{Step $2$:}} We upper bound $C_s^{\rm{Res}}$ by the secrecy capacity when the delivery channel to the adversary is replaced by a discrete memoryless binary erasure channel, with erasure probability $1-\alpha$, denoted as $C_s^{\rm{DM}}$. The proof for this step follows the same lines as in \cite[Section V]{nafea2018new}. The idea is when the binary erasure channel produces a number of erasures greater
than or equal to $(1-\alpha)n$, the adversary's channel in this discrete memoryless setup is worse than its channel in the former model, i.e., when it encounters exactly $(1-\alpha)n$ erasures and is able to select their positions. Hence, $C_s^{\rm{Res}}\leq C_s^{\rm{DM}}$ for this case. The result follows by using Sanov's theorem in
method of types \cite[Theorem 11.4.1]{cover2006elements} to show that the probability of the binary erasure channel causing a number of erasures less than $(1-\alpha)n$ goes to zero as $n \rightarrow\infty$.

{\it{Step $3$:}} From Step $1$, each receiver has a secure cache of size $\frac{n}{2}$ bits. Since increasing the cache sizes cannot decrease the achievable file rate, we upper bound $C_s^{\rm{DM}}$ with the maximum achievable file rate when each receiver has a cache of size $n$ bits, in which it stores $\bX_c^n$. Receiver $j=1,2,$ uses $\bX_c^n$, $\bX_{\bd}^n$ to decode $W_{d_j}$, i.e., $\hat{W}_{d_j}=g_{\bd,j}(\bX_c^n,\bX_{\bd}^n)$. This setup is equivalent to a single receiver, which has a cache of size $n$ bits, demands two files $W_{d_1}$, $W_{d_2}$, and uses the decoder $g_{\bd}\triangleq \{g_{\bd,1},g_{\bd,2}\}$. Let $C_s^{\rm{SR}}$ be the maximum achievable file rate for this single receiver model. We have $C_s^{\rm{DM}}\leq C_s^{\rm{SR}}$. Next, we upper bound $C_s^{\rm{SR}}$.

Let $M_{\rm{D}}$ denote the fraction of the size-$n$ bits cache memory dedicated to store (coded or uncoded) information bits, and let $M_{\rm{K}}$ denote the fraction dedicated to store key bits. That is, $M_{\rm{D}}+M_{\rm{K}}=1$. Let $S_{\rm{D}}$ denote the information bits stored in this memory, i.e., $S_{\rm{D}}=f_{\rm{D}}(W_{[1:D]})$ and $H(S_{\rm{D}})=n M_{\rm{D}}$. We use the following lemma to upper bound $C_s^{\rm{SR}}$.
\begin{lemma}\cite[Lemma 1]{zewail2018wiretap}
For a fixed allocation of $M_{\rm{D}},M_{\rm{K}}$, and a receiver who demands the files $W_{d_1},W_{d_2}$, the secrecy rate for the single receiver model is upper bounded as 
\begin{align}
\label{eq:UB_1}
2R_s^{\rm{SR}}\leq \min\{1,1-\alpha+M_{\rm{K}}\}+\frac{1}{n} I(W_{d_1},W_{d_2};S_{\rm{D}}).
\end{align}
\end{lemma}

Notice that (\ref{eq:UB_1}) holds for any demand pair $\bd=(d_1,d_2)$ such that $d_1\neq d_2$, i.e., the worst-case demands. Summing over all such demands, we have
\begin{align}
\label{eq:UB_2}
\nonumber &2R_s^{\rm{SR}}\leq \min\{1,1-\alpha+M_{\rm{K}}\}\\
&\; +\frac{1}{nD(D-1)} \sum_{d_1,d_2\in[1:D],\;d_1\neq d_2} I(W_{d_1},W_{d_2};S_{\rm{D}}).
\end{align}
The second term on the RHS of (\ref{eq:UB_2}) can be written as
\begin{align}
\label{eq:UB_3_1}
\nonumber &\frac{1}{nD(D-1)} \sum_{d_1,d_2\in[1:D],\;d_1\neq d_2} I(W_{d_1},W_{d_2};S_{\rm{D}})\\
\nonumber &=\frac{1}{nD} \sum_{d_1\in[1:D]}\;I(W_{d_1};S_{\rm{D}})\\
&\qquad +\frac{1}{nD(D-1)}\sum_{d_1,d_2\in[1:D],\; d_1\neq d_2}\;I(W_{d_2};S_{\rm{D}}|W_{d_1})\\
\label{eq:UB_3}
\nonumber &\leq \frac{1}{nD} \sum_{d_1\in[1:D]}\;I(W_{d_1};S_{\rm{D}})\\
&\;\; +\frac{1}{nD(D-1)}\sum_{d_1\in[1:D]}(\sum_{d_2\in[1:D]}\;I(W_{d_2};S_{\rm{D}}|W_{d_1})).
\end{align}

For any $d_1\in[1:D]$, we have 
\begin{align}
\label{eq:UB_4}
\nonumber \sum_{d_2\in[1:D]}&\;I(W_{d_2};S_{\rm{D}}\big|W_{d_1})\\
&=\sum_{d_2=1}^D\;[H(W_{d_2}\big|W_{d_1})-H(W_{d_2}\big|W_{d_1},S_{\rm{D}})]\\
\label{eq:UB_5}
\nonumber &\leq \sum_{d_2=1}^D\;[H(W_{d_2}\big|W_1,W_2,\cdots,W_{d_2-1},W_{d_1})\\
&\;\;\;-H(W_{d_2}\big|W_1,W_2,\cdots,W_{d_2-1},W_{d_1},S_{\rm{D}})]\\
\label{eq:UB_6}
&=I(W_1,W_2,\cdots,W_{\rm{D}};S_{\rm{D}}\big|W_{d_1})\\
\label{eq:UB_7}
&\leq H(S_{\rm{D}})=nM_{\rm{D}},
\end{align}
where (\ref{eq:UB_5}) follows because when $d_2=d_1$, $H(W_{d_2}|W_{d_1})=H(W_{d_2}|W_1,W_2,\cdots,W_{d_2-1},W_{d_1})=0$, and when $d_2\neq d_1$, $H(W_{d_2}|W_{d_1})=H(W_{d_2}|W_1,W_2,\cdots,W_{d_2-1},W_{d_1})=H(W_{d_2})$. Similarly, we have 
\begin{align}
\label{eq:UB_8}
&\sum_{d_1\in[1:D]}\;I(W_{d_1};S_{\rm{D}})\leq H(S_{\rm{D}})=nM_{\rm{D}}.
\end{align}

Substituting (\ref{eq:UB_7}) and (\ref{eq:UB_8}) in (\ref{eq:UB_3}) gives
\begin{align}
\label{eq:UB_9}
&\frac{1}{nD(D-1)} \underset{\begin{subarray}{c}d_1,d_2\in[1:D]\\ d_1\neq d_2\end{subarray}} \sum I(W_{d_1},W_{d_2};S_{\rm{D}})\leq \frac{2D-1}{D(D-1)}M_{\rm{D}}.
\end{align} 

Thus, using (\ref{eq:UB_2}) and (\ref{eq:UB_9}), $R_s^{\rm{SR}}$ is further upper bounded as 
\begin{align}
\label{eq:UB_10}
R_s^{\rm{SR}}\leq \frac{1}{2}[\min\{1,1-\alpha+M_{\rm{K}}\}+\frac{2D-1}{D(D-1)}M_{\rm{D}}].
\end{align}

Finally, by maximizing over all possible allocations for $M_{\rm{D}}$, $M_{\rm{K}}$ such that $M_{\rm{D}}+M_{\rm{K}}=1$, we get 
\begin{align}
\label{eq:UB_11}
C_s^{\rm{SR}}&\leq \frac{1}{2}\underset{\begin{subarray}{c}M_{\rm{D}},M_{\rm{K}}:\\ M_{\rm{D}}+M_{\rm{K}}=1\end{subarray}}\max\{\min\{1,1-\alpha+M_{\rm{K}}\}+\frac{2D-1}{D(D-1)}M_{\rm{D}}\}\\
\label{eq:UB_12}
&=\frac{1}{2}[1+\frac{2D-1}{D(D-1)}(1-\alpha)].
\end{align}
Equation (\ref{eq:UB_12}) follows because, for $D\geq 3$, the maximum occurs at $M_{\rm{K}}=\alpha$ and $M_{\rm{D}}=1-\alpha$. This completes the proof for Theorem \ref{thm:Thm3}. 

\begin{remark}
An upper bound considering uncoded placement only can be derived as follows. The same analysis as in (\ref{eq:UB_1})-(\ref{eq:UB_12}) carries through with $I(W_{d_2};S_{\rm{D}}|W_{d_1})$ in (\ref{eq:UB_3_1}) is equal to $I(W_{d_2};S_{\rm{D}})$. Hence the right hand side of (\ref{eq:UB_9}) is replaced by $\frac{2M_{\rm{D}}}{D}$. The resulting bound $R_s\leq \frac{1}{2}+\frac{(1-\alpha)}{D}$ is tighter than (\ref{eq:UB_12}).
\end{remark}

\section{Discussion}\label{Discussion}
\subsection{Variable Cache Memories}
While the fixed-size cache memory setup considered in this work can be seen as a clean basic model for the intricate problem in consideration, it also allows us to obtain results and insights that are generalizable to more involved cache memory models. The extension to variable memory sizes can be done by considering multiple communication blocks for placement. Our results and coding scheme readily apply to an adversary model whose tapping capability during delivery is normalized with respect to tapping during placement; $\mu_1+B\mu_2\leq \mu$, $B$ is the number of communication blocks for placement. This is a reasonable assumption given that cache placement generally takes place in a longer period than delivery. The problem turns to be more challenging when the adversary optimizes its tapping uniformly over the multiple blocks for cache placement as well as the delivery phase. This is left for future investigation.

\subsection{Extension to More than Two Users}
The broadcast wiretap channel II in \cite{nafea2017new1} can be generalized to more than two users. In \cite{nafea2017new1}, the achievability scheme does not depend on the number of receivers, and the converse proof can be extended to any broadcast setting of known secrecy capacity, see for example \cite{ekrem2009secrecy,benammar2015secrecy}. It follows that the results in this work can also be generalized to more than two users. In particular, the key enabler for our achievability scheme is the proposed channel coding structure which involves using security embedding code along with the reversed embedding order across the two communication phases. By carefully choosing the messages and encryption keys to be transmitted over cache placement and delivery, the same channel coding structure can be applied to the case of more than two users.  

\subsection{The Broadcast Channel Model for Cache Placement}
It is typical to model cache placement as a noiseless channel since placement is assumed to occur when networks are not congested and their rates are assumed to be large enough. Here however we model the cache placement as a broadcast channel communication. The broadcast model avails a clean and tractable solution without compromising its generalizability. A time division multiple access (TDMA) model for cache placement is a special case by imposing an additional constraint in which each receiver has to decode its desired file using only one half of the transmitted codeword. Additionally, the broadcast model is in line with the network information theory literature and it does not limit cache placement to occur over low rate traffic. With the ever-growing user demands, placement and delivery occurring in less asymmetric network loads is likely to be expected in the near future.

\subsection{Larger Cache Sizes Lead to Simple Achievability}
For a library with two files, if the receivers were to have cache memories of size $n$ bits in which they store the transmitted signal during placement, the strong secrecy file rate in Theorem \ref{thm:Thm1} is achievable using a simple wiretap code: The transmitter encodes $W=(W_1,W_2)\in[1:2^{n2R}]$ into a length-$2n$ binary codeword using a wiretap code, and sends the first $n$ bits of this codeword during cache placement and the last $n$ bits during delivery. Each receiver can thus decode both files, and the secrecy of $W_1$ and $W_2$ against the adversary follows by the results in \cite{goldfeld2015semantic,nafea2018new}. In caching problems, the relevant setup however is when the receivers have cache memories of limited size with respect to the overall transmission during cache placement. This calls for the limited size cache memory model considered in this paper, which in turn necessitates the use of the more elaborate coding scheme in Section \ref{Proof_Thm1}. 

\section{Conclusion}\label{Con}
We have introduced the caching broadcast channel with {\it{a wire and cache}} tapping adversary of type II. Each receiver is equipped with a fixed-size cache memory, and the adversary is able to tap into a subset of its choice of the transmitted symbols during cache placement, delivery, or both. The legitimate terminals have no knowledge about the fractions of the tapped symbols in each phase, nor their positions. Only the size of the overall tapped set is known. We have identified the strong secrecy capacity of this model-- the maximum achievable file rate while keeping the overall library secure-- when the transmitter's library has two files. We have derived lower and upper bounds for the strong secrecy file rate when the transmitter has more than two files in its library. We have devised an achievability scheme which combines wiretap coding, security embedding codes, one-time pad keys, and coded caching techniques. 

The results presented in this paper highlight the robustness of (stochastic) coding in a cache-aided network, against a smart adversary who is able to perform a strategic attack jointly optimized over cache placement and delivery phases. Future directions that can build on this work include  exploring variable cache memory sizes, models with more than two end-users, other network topologies, and models with noisy legitimate channels. 

%\section*{Acknowledgment}
%The authors would like to thank Prof. Abbas El Gamal for useful discussions about the modeling details considered in this work, and for the suggestion to develop the results building on the simpler cases.
%%------------------------------------------------------------------------------------------------

\appendices
\section{Secrecy Constraint for Setting $1$}\label{AppendixA}
For every $S_1\subseteq [1:n]$ satisfying $|S_1|=\mu$, we have 
\begin{align}
\label{eq:AppendixA_1}
\nonumber &\limitn I(W_1,W_2;\bZ_{S_1}^n)\\
&=\limitn I(W_1^{(1)}, W_1^{(2)},W_2^{(1)},W_2^{(2)},W_{1,s},W_{2,s};\bZ_{S_1}^n)\\
\label{eq:AppendixA_2}
&=\limitn I (W_1^{(1)}, W_1^{(2)},W_2^{(1)},W_2^{(2)};\bZ_{S_1}^n )\\
\label{eq:AppendixA_3}
&\leq \limitn I(W_1^{(1)}\oplus W_2^{(1)},W_1^{(2)}\oplus W_2^{(2)};\bZ_{S_1}^n)\\
\label{eq:AppendixA_4}
&=\limitn I (M_c;\bZ_{S_1}^n)=0.
\end{align}
The adversary's observation over cache placement, $\bZ_{S_1}^n$, results from sending $M_c=\{M_{c,1},M_{c,2}\}$; $M_{c,1}=W_1^{(1)}\oplus W_2^{(1)}$. $M_{c,2}=W_1^{(2)}\oplus W_2^{(2)}$. Thus, (\ref{eq:AppendixA_2}) follows since $\bZ_{S_1}^n$ does not depend on $\{W_{1,s},W_{2,s}\}$ and  (\ref{eq:AppendixA_3}) follows due to the Markov chain $\{W_1^{(1)}, W_1^{(2)},W_2^{(1)},W_2^{(2)}\}-\{W_1^{(1)}\oplus W_2^{(1)},W_1^{(2)}\oplus W_2^{(2)}\}-\bZ_{S_1}^n$. The second equality in (\ref{eq:AppendixA_4}) follows from \cite[Theorem 2]{goldfeld2015semantic}, and since the rate of $M_c$ is less than $1-\alpha$. 

\section{Secrecy Constraint for Setting $2$}\label{AppendixB}
For every $S_2\subseteq [1:n]$ satisfying $|S_2|=\mu$ and any $\bd=(d_1,d_2)$, $d_1,d_2\in\{1,2\}$, 
\begin{align}
\label{AppendixB_1}
&I(W_1,W_2;\bZ_{S_2}^n)=I(W_{d_1}^{(2)},W_{d_2}^{(1)},W_{d_1,s}, W_{d_2,s};\bZ_{S_2}^n)\\
\label{AppendixB_2}
&=I(W_{d_1,s}, W_{d_2,s};\bZ_{S_2}^n\big|W_{d_1}^{(2)},W_{d_2}^{(1)})+I(W_{d_1}^{(2)},W_{d_2}^{(1)};\bZ_{S_2}^n)\\
\label{AppendixB_3}
\nonumber &\leq I(W_{d_1,s}, W_{d_2,s};W_{d_1,s}\oplus K_1, W_{d_2,s}\oplus K_2\big|W_{d_1}^{(2)},W_{d_2}^{(1)})\\
&\qquad\qquad+ I(W_{d_1}^{(2)},W_{d_2}^{(1)};\bZ_{S_2}^n)\\
\label{AppendixB_4}
\nonumber &=I(W_{d_1,s},W_{d_2,s};W_{d_1,s}\oplus K_1,W_{d_2,s}\oplus K_2)\\
& \qquad\qquad + I(W_{d_1}^{(2)},W_{d_2}^{(1)};\bZ_{S_2}^n)\\
\label{AppendixB_5}
&=I(W_{d_1}^{(2)},W_{d_2}^{(1)};\bZ_{S_2}^n)\\
\label{AppendixB_6}
&=I(M_{\bd};\bZ_{S_2}^n).
\end{align}
The adversary's observation over delivery, $\bZ_{S_2}^n$, results from sending $M_{\bd}=\{W_{d_1}^{(2)},W_{d_2}^{(1)}\}$ and $\tilde{M}_{\bd}=\{W_{d_1,s}\oplus K_1,W_{d_2,s}\oplus K_2\}$. (\ref{AppendixB_1}) follows because $\bZ_{S_2}^n$ depends only on $W_{d_1}^{(2)}$, $W_{d_2}^{(1)}$, $W_{d_1,s}$, $W_{d_2,s}$. (\ref{AppendixB_3}) follows from the Markov chain $\{W_{d_1,s},W_{d_2,s}\}-\{W_{d_1}^{(2)},W_{d_2}^{(1)},W_{d_1,s}\oplus K_1,W_{d_2,s}\oplus K_2\}-\bZ_{S_2}^n$. (\ref{AppendixB_4}) follows because $\{W_{d_1}^{(2)},W_{d_2}^{(1)}\}$, $\{W_{d_1,s},W_{d_2,s},K_1,K_2\}$ are independent. The randomization message for the wiretap code during delivery  $\tilde{M}_{\bd}$ is independent from the message $M_{\bd}$. Using (\ref{AppendixB_6}), \cite[Theorem 2]{goldfeld2015semantic},
\begin{align}
\label{AppendixB_7}
\nonumber &\limitn\; \max_{S_2\subseteq [1:n]: |S_2|=\mu}\;I(W_1,W_2;\bZ_{S_2}^n)\\
&\qquad \qquad \leq \limitn \max_{S_2\subseteq [1:n]:\;|S_2|=\mu}\;I(M_{\bd};\bZ_{S_2}^n)=0.
\end{align}

\section{Secrecy Constraint for Setting $3$ when $\alpha_1\geq \alpha_2$}\label{AppendixC}
For a fixed choice of $S_1,S_2\subseteq [1:n]$ s.t. $|S_1|+|S_2|=\mu$, and any $\bd=(d_1,d_2)$, $d_1,d_2\in\{1,2\}$, 
\begin{align}
\label{AppendixC_1}
\nonumber &I(W_1,W_2;\bZ_{S_1}^n,\bZ_{S_2}^n)\\
&=I(W_1^{(1)},W_1^{(2)},W_2^{(1)},W_2^{(2)},W_{1,s},W_{2,s};\bZ_{S_1}^n,\bZ_{S_2}^n)\\
\label{AppendixC_2}
\nonumber &=I(W_1^{(1)}\oplus W_2^{(1)},W_1^{(2)}\oplus W_2^{(2)},W_{d_1}^{(2)},W_{d_2}^{(1)},\\
&\qquad \qquad W_{d_1,s},W_{d_2,s};\bZ_{S_1}^n,\bZ_{S_2}^n)\\
\label{AppendixC_3}
&=I(M_c,M_{\bd};\bZ_{S_1}^n,\bZ_{S_2}^n)\\
\label{AppendixC_4}
&=I(M_c;\bZ_{S_1}^n,\bZ_{S_2}^n)+I(M_{\bd};\bZ_{S_1}^n,\bZ_{S_2}^n\big|M_c)\\
\label{AppendixC_5}
\nonumber &=I(M_c;\bZ_{S_1}^n)+I(M_c;\bZ_{S_2}^n\big|\bZ_{S_1}^n)+I(M_{\bd};\bZ_{S_2}^n\big|M_c)\\
&\qquad \qquad \qquad +I(M_{\bd};\bZ_{S_1}^n\big|M_c,\bZ_{S_2}^n).
\end{align}
(\ref{AppendixC_2}) follows because, for any $d_1,d_2\in\{1,2\}$, there is a bijective map between $\{W_1^{(1)},W_1^{(2)},W_2^{(1)},W_2^{(2)}\}$ and $\{W_1^{(1)}\oplus W_2^{(1)},W_1^{(2)}\oplus W_2^{(2)},W_{d_1}^{(2)},W_{d_2}^{(1)}\}$. 

From (\ref{eq:example1_1}), (\ref{eq:example1_2}); $M_c$ and $M_{\bd}$ are independent. $\bZ_{S_1}^n$ results from sending $M_c$, while $\bZ_{S_2}^n$ results from sending $M_{\bd}$. For a fixed choice of $S_1,S_2$, $\{M_c,\bZ_{S_1}^n\}$ are independent from $\bZ_{S_2}^n$. Thus, we have 
\begin{align}
\label{AppendixC_6}
I(M_c;\bZ_{S_2}^n|\bZ_{S_1}^n)=0.
\end{align}
In addition, $\{M_{\bd},\bZ_{S_2}^n\}$ are independent from $M_c$. Thus,
\begin{align}
\label{AppendixC_7}
I(M_{\bd};\bZ_{S_2}^n|M_c)&=H(\bZ_{S_2}^n|M_c)-H(\bZ_{S_2}^n|M_c,M_{\bd})\\
\label{AppendixC_8}
&=H(\bZ_{S_2}^n|M_c)-H(\bZ_{S_2}^n|M_{\bd})\\
\label{AppendixC_9}
&\leq I(M_{\bd};\bZ_{S_2}^n).
\end{align}
Finally, using the Markov chain $\{M_{\bd},\bZ_{S_2}^n\}-M_c-\bZ_{S_1}^n$, 
\begin{align}
\label{AppendixC_10}
\nonumber & I(M_{\bd};\bZ_{S_1}^n|M_c,\bZ_{S_2}^n)\\
&\qquad =H(\bZ_{S_1}^n|M_c,\bZ_{S_2}^n)-H(\bZ_{S_1}^n|M_c,\bZ_{S_2}^n,M_{\bd})\\
\label{AppendixC_11}
&\qquad \leq H(\bZ_{S_1}^n)-H(\bZ_{S_1}^n|M_c)=I(M_c;\bZ_{S_1}^n).
\end{align}

Substituting (\ref{AppendixC_6}), (\ref{AppendixC_9}), and (\ref{AppendixC_11}) in (\ref{AppendixC_5}), 
\begin{align}
\label{AppendixC_12}
I(W_1,W_2;\bZ_{S_1}^n,\bZ_{S_2}^n)\leq 2I(M_c;\bZ_{S_1}^n)+I(M_{\bd};\bZ_{S_2}^n).
\end{align}

The rates of $M_c$, $M_{\bd}$ are $1-\alpha_1-\epsilon_n$, $1-\alpha_2-\epsilon_n$, respectively. By applying \cite[Theorem 2]{goldfeld2015semantic} to (\ref{AppendixC_12}), 
\begin{align}
\nonumber &\limitn \;\underset{\begin{subarray}{c} S_1,S_2\subseteq [1:n]:\\|S_1|+|S_2|=\mu\end{subarray}}\max I(W_1,W_2;\bZ_{S_1}^n,\bZ_{S_2}^n)\\
\label{AppendixC_13}
\nonumber &\qquad \leq 2 \limitn\;\underset{S_1\subseteq[1:n]:\;|S_1|=\mu_1}\max\;I(M_c;\bZ_{S_1}^n)\\
&\qquad\;+\limitn\;\underset{S_2\subseteq[1:n]:\;|S_2|=\mu_2}\max\;I(M_{\bd};\bZ_{S_2}^n)=0.
\end{align}

\section{Secrecy Constraint for Setting $3$ when $\alpha_1< \alpha_2$}\label{AppendixD}
For notational simplicity, let us define 
\begin{align}
\label{AppendixD_0_1}
&M_{c,1\setminus K_1}=W_1^{(1)}\oplus W_2^{(1)},\; M_{c,2 \setminus K_2}=W_1^{(2)}\oplus W_2^{(2)}\\
\label{AppendixD_0_2}
&M_{c\setminus K}=\{M_{c,1\setminus K_1},M_{c,2\setminus K_2}\}.
\end{align}

For fixed $S_1,S_2\subseteq [1:n]$ such that $|S_1|+|S_2|=\mu$, and any $\bd=(d_1,d_2)$, $d_1,d_2\in\{1,2\}$, 
\begin{align}
\label{AppendixD_1}
\nonumber & I(W_1,W_2;\bZ_{S_1}^n,\bZ_{S_2}^n)\\
\nonumber &=I(W_1^{(1)}\oplus W_2^{(1)},W_1^{(2)}\oplus W_2^{(2)},W_{d_1}^{(2)},W_{d_2}^{(1)},\\
& \qquad \qquad \qquad \qquad  W_{d_1,s},W_{d_2,s};\bZ_{S_1}^n,\bZ_{S_2}^n)\\
\label{AppendixD_2}
&=I(M_{c\setminus K},M_{\bd},W_{d_1,s},W_{d_2,s};\bZ_{S_1}^n,\bZ_{S_2}^n)\\
\label{AppendixD_3}
\nonumber &=I(M_{c\setminus K};\bZ_{S_1}^n,\bZ_{S_2}^n)+I(M_{\bd};\bZ_{S_1}^n,\bZ_{S_2}^n|M_{c\setminus K})\\
&\qquad \qquad +I(W_{d_1,s},W_{d_2,s};\bZ_{S_1}^n,\bZ_{S_2}^n|M_{\bd},M_{c\setminus K}).
\end{align}

From (\ref{eq:example2_2}), (\ref{eq:example2_3}), $M_c$ is independent from $\{M_{\bd},\tilde{M}_{\bd}\}$. $\bZ_{S_1}^n$ results from sending $M_{c}=\{M_{c\setminus K}, K_1,K_2\}$, and $\bZ_{S_2}^n$ results from sending $M_{\bd}=\{W_{d_1}^{(2)},W_{d_2}^{(1)}\}$, $\tilde{M}_{\bd}=\{W_{d_1,s}\oplus K_1, W_{d_2,s}\oplus W_2\}$. 

We upper bound each term on the RHS of (\ref{AppendixD_3}). For the third term, we have
\begin{align}
\label{AppendixD_4}
\nonumber & I(W_{d_1,s},W_{d_2,s};\bZ_{S_1}^n,\bZ_{S_2}^n\big|M_{\bd},M_{c\setminus K})\\
&\leq I(W_{d_1,s},W_{d_2,s};\tilde{M}_{\bd}\big|M_{\bd},M_{c\setminus K})\\
\label{AppendixD_5}
&=I(W_{d_1,s},W_{d_2,s};W_{d_1,s}\oplus K_1,W_{d_2,s}\oplus K_2)=0,
\end{align}
where (\ref{AppendixD_4}) follows due to the Markov chain $\{W_{d_1,s},W_{d_2,s}\}-\{M_{c\setminus K},M_{\bd},\tilde{M}_{\bd}\}-\{\bZ_{S_1}^n,\bZ_{S_2}^n\}$, and (\ref{AppendixD_5}) follows because $\tilde{M}_{\bd}$ is independent from $\{W_{d_1,s},W_{d_2,s},M_{\bd},M_{c\setminus K}\}$.

For fixed $S_1,S_2$, $\bZ_{S_2}^n$ is independent from $\{M_c, \bZ_{S_1}^n\}$. Thus, the first term is bounded as 
\begin{align}
\label{AppendixD_6}
I&(M_{c\setminus K};\bZ_{S_1}^n,\bZ_{S_2}^n)\leq I(M_{c};\bZ_{S_1}^n,\bZ_{S_2}^n)\\
\label{AppendixD_7}
&=I(M_{c};\bZ_{S_1}^n)+I(M_{c};\bZ_{S_2}^n\big|\bZ_{S_1}^n)= I(M_{c};\bZ_{S_1}^n).
\end{align}

For the second term on the RHS of (\ref{AppendixD_3}), we have
\begin{align}
\label{AppendixD_9}
\nonumber & I(M_{\bd};\bZ_{S_1}^n,\bZ_{S_2}^n\big|M_{c\setminus K})\\
&=I(M_{\bd};\bZ_{S_2}^n\big|M_{c\setminus K})+I(M_{\bd};\bZ_{S_1}^n\big|M_{c\setminus K},\bZ_{S_2}^n).
\end{align}
Notice that $M_{c\setminus K}$ and $\bZ_{S_2}^n$ are conditionally independent given $M_{\bd}$. Thus,
\begin{align}
\label{AppendixD_10}
\nonumber I(M_{\bd};\bZ_{S_2}^n\big|M_{c\setminus K})&=H(\bZ_{S_2}^n\big|M_{c \setminus K})-H(\bZ_{S_2}^n\big|M_{\bd})\\
&\leq I(M_{\bd};\bZ_{S_2}^n).
\end{align}
In addition, using the independence between $\{M_{\bd},\bZ_{S_2}^n\}$ and $\{M_c,\bZ_{S_1}^n\}$, we have
\begin{align}
\label{AppendixD_11}
\nonumber &I(M_{\bd};\bZ_{S_1}^n\big|M_{c\setminus K},\bZ_{S_2}^n)\\
&=H(\bZ_{S_1}^n\big|M_{c\setminus K},\bZ_{S_2}^n)-H(\bZ_{S_1}^n\big|M_{c\setminus K},M_{\bd},\bZ_{S_2}^n)\\
\label{AppendixD_12}
&\leq H(\bZ_{S_1}^n)-H(\bZ_{S_1}^n\big|M_{c\setminus K},K_1,K_2,M_{\bd},\bZ_{S_2}^n)\\
\label{AppendixD_13}
&= H(\bZ_{S_1}^n)-H(\bZ_{S_1}^n\big|M_{c})=I(M_c;\bZ_{S_1}^n).
\end{align}

Substituting (\ref{AppendixD_10}) and (\ref{AppendixD_13}) in (\ref{AppendixD_9}) gives
\begin{align}
\label{AppendixD_14}
I(M_{\bd};\bZ_{S_1}^n,\bZ_{S_2}^n\big|M_{c\setminus K})\leq I(M_{\bd};\bZ_{S_2}^n)+I(M_c;\bZ_{S_1}^n).
\end{align}

Finally, substituting (\ref{AppendixD_5}), (\ref{AppendixD_7}), (\ref{AppendixD_14}) in (\ref{AppendixD_3}), and applying \cite[Theorem 2]{goldfeld2015semantic}, we have  
\begin{align}
\label{AppendixD_15}
\limitn &\;\underset{\begin{subarray}{c} S_1,S_2\subseteq [1:n]:\\|S_1|+|S_2|=\mu\end{subarray}}\max I(W_1,W_2;\bZ_{S_1}^n,\bZ_{S_2}^n)=0,
\end{align}
since the rates of $M_c$ and $M_{\bd}$ are $1-\alpha_1-\epsilon_n$ and $1-\alpha_2-\epsilon_n$, respectively.
  
\section{Secrecy Constraint for Setting $4$}\label{AppendixE}
From(\ref{eq:M_c})--(\ref{eq:M_d_tilde}), $M_c$ is independent from $\tilde{M}_c$;  $M_{\bd}$ is independent from $\tilde{M}_{\bd}$, and $\{M_c,\tilde{M}_c\}$ are independent from $\{M_{\bd},\tilde{M}_{\bd}\}$. Conditioned on a fixed choice of $S_1,S_2,$ satisfying $\{|S_1|=\mu,\; |S_2|=0\}$ or $\{|S_1|=0,\;|S_2|=\mu\}$, define the random variable
\begin{align}
\label{AppendixE_1}
\bZ_S^n\triangleq \bZ_{S_1}^n\;\mathbbm{1}_{\{|S_2|=0\}}+\bZ_{S_2}^n \;\mathbbm{1}_{\{|S_1|=0\}}.
\end{align}
Note that $\bZ_S^n$ only has a well-defined probability distribution when conditioned on the event $\{S_1,S_2\}$, since a prior distribution on these subsets is not defined. For this fixed choice of the subsets, and any $\bd=(d_1,d_2)$, $d_1,d_2\in\{1,2\}$, we have
\begin{align}
\label{AppendixE_3}
\nonumber & I\big(W_1,W_2;\bZ_{S_1}^n,\bZ_{S_2}^n\big)
\\
\nonumber &=I(W_1^{(1)}\oplus W_2^{(1)},W_1^{(2)}\oplus W_2^{(2)},W_{d_1}^{(2)},W_{d_2}^{(1)},\\
&\qquad \qquad \qquad \qquad W_{d_1,s},W_{d_2,s};\bZ_{S_1}^n,\bZ_{S_2}^n)\\
\label{AppendixE_4}
&=I(M_c,M_{\bd},W_{d_1,s},W_{d_2,s};\bZ_S^n)\\
\label{AppendixE_5}
\nonumber &=\mathbbm{1}_{\{|S_2|=0\}}\;I(M_c,M_{\bd},W_{d_1,s},W_{d_2,s};\bZ_S^n\Big|\{|S_2|=0\})\\
&\qquad +\mathbbm{1}_{\{|S_1|=0\}}\;I(M_c,M_{\bd},W_{d_1,s},W_{d_2,s};\bZ_S^n\Big|\{|S_1|=0\})\\
\label{AppendixE_6}
\nonumber &=\mathbbm{1}_{\{|S_2|=0\}}\;I(M_c,M_{\bd},W_{d_1,s},W_{d_2,s};\bZ_{S_1}^n)\\
&\qquad +\mathbbm{1}_{\{|S_1|=0\}}\;I(M_c,M_{\bd},W_{d_1,s},W_{d_2,s};\bZ_{S_2}^n)\\
\label{AppendixE_7}
\nonumber &=\mathbbm{1}_{\{|S_2|=0\}}\;I(M_c;\bZ_{S_1}^n)\\
& \qquad \qquad +\mathbbm{1}_{\{|S_1|=0\}}\;I(M_{\bd},W_{d_1,s},W_{d_2,s};\bZ_{S_2}^n)\\
\label{AppendixE_8}
&\leq \mathbbm{1}_{\{|S_2|=0\}}\;I(M_c;\bZ_{S_1}^n)+\mathbbm{1}_{\{|S_1|=0\}}\;I(M_{\bd};\bZ_{S_2}^n).
\end{align}
(\ref{AppendixE_7}) follows because (i) $\bZ_{S_1}^n$ results from $\{M_c,\tilde{M}_c\}$ which are independent from $\{M_{\bd},W_{d_1,s},W_{d_2,s}\}$, and (ii) $\bZ_{S_2}^n$ is conditionally independent from $M_c$ given $\{M_{\bd},W_{d_1,s},W_{d_2,s}\}$, due to the Markov chain $M_c-\{M_{\bd},W_{d_1,s},W_{d_2,s}\}-\{M_{\bd},\tilde{M}_{\bd}\}-\bZ_{S_2}^n$. (\ref{AppendixE_8}) follows using the same steps in (\ref{AppendixB_1})--(\ref{AppendixB_6}).

Finally, since $\tilde{M}_c$ is independent from $M_c$; $\tilde{M}_{\bd}$ is independent from $M_{\bd}$, and the rates of $M_c$, $\tilde{M}_{\bd}$ are both equal to $1-\alpha-\epsilon_n$, we have 
\begin{align}
\label{AppendixE_9}
\nonumber &\limitn\;\underset{\begin{subarray}{c} S_1,S_2\subseteq [1:n]:\\|S_1|+|S_2|=\mu\end{subarray}}\max I(W_1,W_2;\bZ_{S_1}^n,\bZ_{S_2}^n)\\
&\;\; =\limitn \;\underset{\begin{subarray}{c} S_1,S_2\subseteq [1:n]:\\|S_i|=0,\;|S_j|=\mu\\i,j\in\{1,2\},\;i\neq j\end{subarray}}\max I(W_1,W_2;\bZ_{S_1}^n,\bZ_{S_2}^n)\\
\label{AppendixE_10}
\nonumber & \leq \limitn\;\max\big\{\underset{S_1\subseteq[1:n]:\;|S_1|=\mu}\max\;I(M_c;\bZ_{S_1}^n),\\
&\qquad \qquad \qquad \underset{S_2\subseteq[1:n]:\;|S_2|=\mu}\max\;I(M_{\bd};\bZ_{S_2}^n)\big\}\\
\label{AppendixE_11}
\nonumber &=\max\big\{\limitn\;\;\underset{S_1\subseteq[1:n]:\;|S_1|=\mu}\max\;I(M_c;\bZ_{S_1}^n),\\
&\qquad\qquad \limitn\;\;\underset{S_2\subseteq[1:n]:\;|S_2|=\mu}\max\;I(M_{\bd};\bZ_{S_2}^n)\big\}=0,
\end{align}
(\ref{AppendixE_10}) follows from (\ref{AppendixE_8}); (\ref{AppendixE_11}) follows since both limits exist and equal to zero, using \cite[Theorem 2]{goldfeld2015semantic}.

\bibliographystyle{IEEEtran}
\bibliography{MyLib}

\begin{IEEEbiography}[{\includegraphics[width=1in,height=1.25in,keepaspectratio]{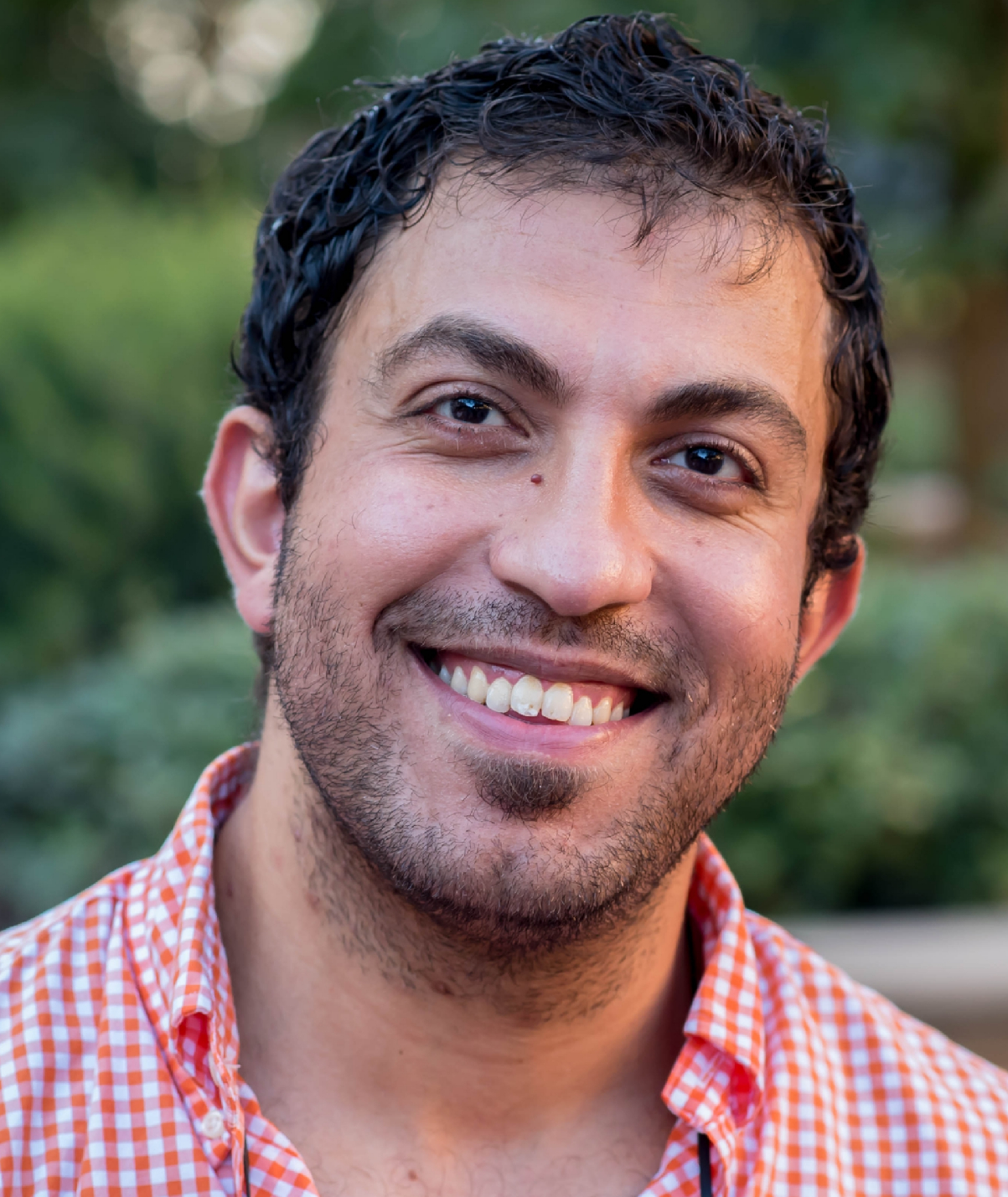}}]{Mohamed Nafea} (Member, IEEE) received the B.Sc. degree in electrical engineering from Alexandria University, Egypt, in 2010; the M.Sc. degree in wireless communications from Nile University, Egypt, in 2012; the M.A. degree in mathematics, and the Ph.D. degree in electrical engineering from The Pennsylvania State University, University Park, in 2017 and 2018, respectively. He is an Assistant Professor at the University of Detroit Mercy, Electrical \& Computer Engineering Department, since Fall 2020. He spent a year as a Postdoctoral Scholar at Georgia Institute of Technology, Electrical \& Computer Engineering Department. His research interests include network information theory, information theoretic security and privacy, algorithmic fairness, statistical machine learning, and causal structure learning. 
\end{IEEEbiography}

\begin{IEEEbiography}[{\includegraphics[width=1in,height=1.25in,clip,keepaspectratio]{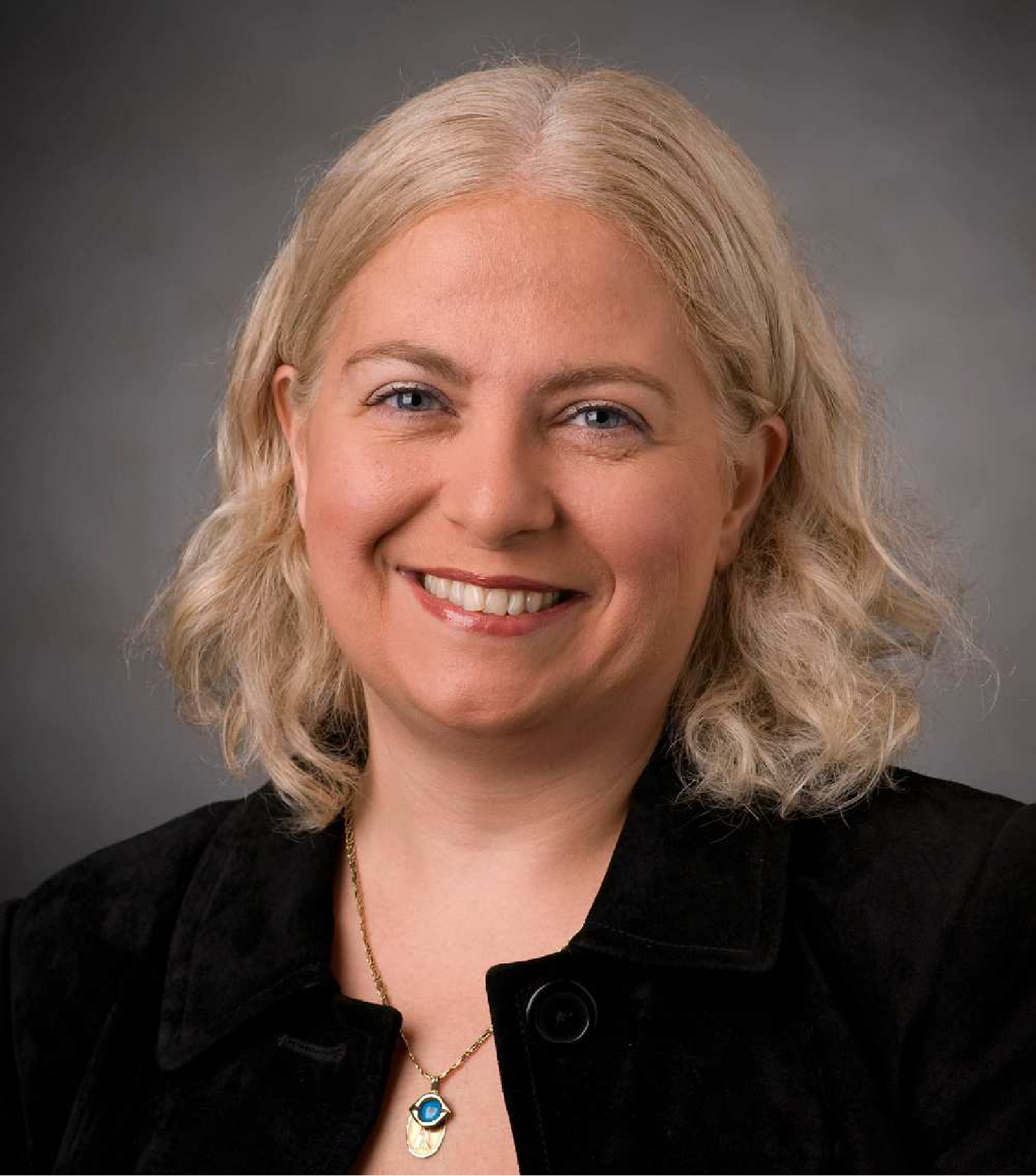}}]
{Aylin Yener} (Fellow, IEEE) received the B.Sc.
degree in electrical and electronics engineering and
the B.Sc. degree in physics from Bogazici University, Istanbul, Turkey, and the M.S. and Ph.D. degrees in electrical and computer engineering from Rutgers University, New Brunswick, NJ, USA. She holds the Roy and Lois Chope Chair in Engineering at The Ohio State University, Columbus Ohio, since 2020, where she is a Professor of electrical and computer engineering, Professor of integrated systems engineering, and Professor of computer science and engineering. Until 2020, she was a University Distinguished Professor of electrical engineering and a Dean’s Fellow at The Pennsylvania State University, University Park, PA, USA, where she joined the faculty as an Assistant Professor in 2002. She was a Visiting Professor of electrical engineering at Stanford University in 2016–2018 and a Visiting Associate Professor in the same department in 2008–2009. Her current research interests are in information security, green communications, caching systems, 6G, and more generally in the fields of information theory, communication theory and networked systems. She received the NSF CAREER Award in 2003, the Best Paper Award in Communication Theory from the IEEE International Conference on Communications in 2010, the Penn State Engineering Alumni Society (PSEAS) Outstanding Research Award in 2010, the IEEE Marconi Prize Paper Award in 2014, the PSEAS Premier Research Award in 2014, the Leonard A. Doggett Award for Outstanding Writing in Electrical Engineering at Penn State in 2014, the IEEE Women in Communications Engineering Outstanding Achievement Award in 2018, the IEEE Communications Society Best Tutorial Paper Award in 2019, and the IEEE Communications Society Communication Theory Technical Achievement Award in 2020. She has been a Distinguished Lecturer for the IEEE Information Theory Society (2019–2021), the IEEE Communications Society (2018–2019) and the IEEE Vehicular Technology Society (2017–2021).

Dr. Yener is currently serving as the Junior Past President of the IEEE Information Theory Society. Previously, she was the President (2020), the Vice President (2019), the Second Vice President (2018), an elected member of the Board of Governors (2015–2018), and the Treasurer (2012–2014) of the IEEE Information Theory Society. She served as the Student Committee Chair for the IEEE Information Theory Society (2007–2011), and was the Co-Founder of the Annual School of Information Theory in North America in 2008. She was a Technical (Co)-Chair for various symposia/tracks at the IEEE ICC, PIMRC, VTC, WCNC, and Asilomar in 2005, 2008–2014 and 2018. Previously, she served as an Editor for IEEE TRANSACTIONS ON COMMUNICATIONS (2009–2012), an Editor for IEEE TRANSACTIONS ON MOBILE COMPUTING (2017–2018), and an Editor and an Editorial Advisory Board Member for IEEE TRANSACTIONS ON WIRELESS COMMUNICATIONS (2001–2012). She also served as a Guest Editor for IEEE TRANSACTIONS ON INFORMATION FORENSICS AND SECURITY in 2011, and IEEE JOURNAL ON SELECTED AREAS IN COMMUNICATIONS in 2015. Currently, she serves as a Senior Editor for IEEE JOURNAL ON SELECTED AREAS IN COMMUNICATIONS and is on the Senior Editorial Board of IEEE JOURNAL ON SELECTED AREAS IN INFORMATION THEORY.
 \end{IEEEbiography}

\end{document}